\documentclass{aa}  
\usepackage{graphicx}

\usepackage{pkg}

\usepackage[pdfborder={0 0 1}, citebordercolor={0 1 0}, linkbordercolor={1 0 0}]{hyperref}

\defcitealias{Benitez-Llambay2020}{BLF20}
\defcitealias{Benitez-Llambay2017}{BL17}

\begin{document} 

\title{Weighing gas-rich starless halos: Dark
matter parameter inference based on gas distributions}
\titlerunning{Weighing gas-rich starless halos}
\authorrunning{F. Turini \and A. Benítez-Llambay}
  \author{
    Francesco Turini\inst{1}
    \and
    Alejandro Benítez-Llambay\inst{2}
    }
    
  \institute{
        Dipartimento di Fisica "Giuseppe Occhialini", Università degli Studi di Milano-Bicocca\\
        \email{f.turini2@campus.unimib.it} 
    \and
        Dipartimento di Fisica "Giuseppe Occhialini", Università degli Studi di Milano-Bicocca\\
        \email{alejandro.benitezllambay@unimib.it} 
    }

   \date{Received Month date, year; accepted Month date, year}

\abstract
{Reionization-limited \ion{H}{I} clouds (RELHICs) are starless dark matter halos that retain a significant neutral hydrogen (\ion{H}{I}) reservoir. The gas resides in near hydrostatic equilibrium within the dark matter potential and in thermal equilibrium with the cosmic ultraviolet background. This simplicity allows analytic frameworks to link observable \ion{H}{I} column densities directly to fundamental dark matter halo structural parameters.}{In this work we systematically assess the accuracy of inferring host halo parameters from RELHIC gas distributions on an object-by-object basis, quantifying biases, intrinsic degeneracies, and the limits of parameter recovery.}{Using RELHICs from a redshift $z=0$ high-resolution cosmological hydrodynamical simulation, we employed Bayesian nested sampling to infer dark matter halo mass and concentration. We evaluated this approach against 3D spherically averaged total gas and \ion{H}{I} density profiles alongside 2D \ion{H}{I} column density profiles.}{While the ensemble inference yields a robust, unbiased recovery of halo virial mass from 3D profiles, individual systems exhibit a mass-concentration degeneracy driven by local environmental density. Overdense environments yield slightly overestimated masses and underestimated concentrations, and underdense regions show the inverse. Consequently, detectable low-mass systems are biased toward higher masses and lower concentrations. Applying this to 2D projected \ion{H}{I} column densities washes out radial features via line-of-sight integration, which preserves mass constraints but exacerbates concentration underestimation.}{The local intergalactic medium modulates the RELHIC gas content, meaning universal boundary conditions force artificial adjustments to the inferred halo parameters to compensate for external pressure. We demonstrate that treating environmental density as a free parameter breaks this degeneracy and completely neutralizes the systematic mass bias. Although concentration recovery remains limited by simulation resolution, the virial mass is exceptionally well constrained, establishing a highly reliable framework for weighing starless halos in upcoming surveys.}


   \maketitle

\section{Introduction}
\label{sec:Introduction}

A cornerstone of the $\Lambda$ cold dark matter ($\Lambda$CDM) cosmological model is hierarchical structure formation ~\citep[][]{Blumenthal1984}. Within this framework, two central, testable predictions emerge for the low-mass dark matter halo population: (i) a steeply rising halo mass function, $dN/dM \propto M^{-1.9}$~\citep{Press1974,Jenkins2001,Angulo2012}, and (ii) a universal, cuspy density profile for virialized halos~\citep{Navarro1996a,Navarro1997}, characterized by a mass-dependent concentration parameter, $c(M)$~\citep[e.g.,][]{Ludlow2016}. Yet, both of these predictions stand in apparent tension with observations of low-mass galaxies in the nearby universe~\citep[see, e.g.,][for a review]{Bullock2017}.

On the one hand, the observed census of low-mass galaxies has revealed an abundance significantly smaller than that of low-mass halos, a problem that manifests itself around our galaxy as the so-called missing-satellites problem~\citep{Klypin1999, Moore1999}. On the other hand, the inner structure of dark matter halos, as inferred from the kinematics of dwarf galaxies, often appears to have a lower central density than predicted, an issue commonly referred to as the ``cusp-core problem''~\citep[e.g.,][and references therein]{Flores1994, Moore1994, deBlok2008, Oh2011, Oh2015}. These apparent discrepancies can be naturally reconciled within the $\Lambda$CDM model once the physics of galaxy formation is considered.

The observed low abundance of halos is resolved if star formation becomes increasingly inefficient in dark matter halos below a characteristic mass, rendering them ``starless'' or too faint for detection~\citep[e.g.,][]{Bullock2000}. The structural problem is accounted for by considering the impact of baryonic feedback, where energetic processes associated with star formation can dynamically heat the dark matter and transform the primordial cusp into a core~\citep[e.g.,][]{Navarro1996b, Governato2012, Penarrubia2012, DiCintio2014}.

The star formation suppression mechanism is not merely a phenomenological convenience but is physically motivated by the thermal history of the Universe. Specifically, the photoionization and photoheating of the intergalactic medium to $T\sim 10^4$ K by the cosmic ultraviolet background (UVB) is the primary cause~\citep[e.g.,][]{Efstathiou1992, Thoul1996, Quinn1996, Bullock2000}. This process raised the cosmological Jeans mass after the epoch of reionization, providing sufficient pressure support to prevent baryonic gas from collapsing into the shallow potential wells of halos below a critical redshift-dependent mass, $M_{\rm crit}(z)$~\citep[][BLF2020 hereafter]{Benitez-Llambay2020}. At the present epoch, this threshold is estimated to be $M_{\rm crit}(0) \approx 10^{9.7} \, \mathrm{M}_\odot$. Consequently, halos below this mass are expected to largely be devoid of stars if they did not form stars prior to reionization~\citep[e.g.,][]{Blumenthal1984, Hoeft2006, Okamoto2009, Sawala2016,  Benitez-Llambay2020,Garcia-Bethencourt2026} or quiescent and faint if they host stars today~\citep{Benitez-Llambay2021, Pereira-Wilson2023}.

The solution to the structural discrepancy is likewise rooted in baryonic physics, but it applies specifically to halos that successfully formed galaxies. In this scenario, energetic feedback from star formation dramatically alters the halo's structure: Supernova-driven gas outflows gravitationally heat the dark matter, causing it to expand and transforming the primordial central cusp into a flattened core~\citep[e.g.,][]{Pontzen2012}. Although initially controversial~\citep[e.g.,][]{Oman2015, Bose2019}, this cusp-to-core transformation is widely accepted as a possible outcome in cosmological hydrodynamical simulations of dwarf galaxies~\citep[e.g.,][and references therein]{DiCintio2014, Chan2015, Tollet2016, Benitez-Llambay2019}. This implies that the inner structure of star-forming halos is not expected to match the predictions of collisionless dark matter-only simulations. Instead, exhibits a strong dependence with the past and present star formation history of the galaxy~\citep[e.g.,][]{DiCintio2014, Benitez-Llambay2019}. The efficiency of this transformation is, however, sensitive to model assumptions, particularly the star formation and feedback prescriptions, a characteristic that results in disparate outcomes dependent on the details of numerical implementations~\citep{Benitez-Llambay2019, Dutton2019} and thereby precludes robust predictions.

Irrespective of these uncertainties, a powerful corollary emerges: Halos that failed to form stars and are starless today should be immune to this baryon-induced transformation. They must therefore retain their pristine, cuspy density profiles, offering a clean test of the fundamental tenets of the $\Lambda$CDM model and the nature of dark matter itself.

Despite their inability to form stars, cosmological simulations and theoretical models predict that the most massive of these starless halos ($M_{200} \gtrsim 2\times 10^9\, \mathrm{M}_\odot$) can nevertheless retain a significant neutral hydrogen (\ion{H}{I}) reservoir, potentially detectable via its 21-cm line emission~\citep[][BL17 hereafter]{Benitez-Llambay2017}. These systems, termed reionization-limited~\ion{H}{I} clouds (RELHICs), are characterized by a remarkable physical simplicity: Their gas is expected to be in near hydrostatic equilibrium within the dark matter halo's gravitational potential and in thermal equilibrium with the cosmic UVB. This simplicity enabled~\citetalias{Benitez-Llambay2017} to develop a predictive analytic model that directly links the fundamental properties of a dark matter halo---namely its virial\footnote{We define virial quantities as those measured within a sphere of mean density equal to 200 times the critical density of the Universe.} mass and concentration---to its total gas and observable~\ion{H}{I} column density profile. Consequently, the model can be used to infer the structural parameters of the host halo from~\ion{H}{i} observations, offering a compelling empirical estimator for the halo mass and concentration in a regime that is otherwise inaccessible to traditional methods.

Although initially applied to zoom-in cosmological simulations, this analytic framework has been further validated by large-volume cosmological hydrodynamical simulations. For instance,~\citetalias{Benitez-Llambay2020} showed that the simple RELHIC model reproduces the ensemble-averaged virial gas mass and size of simulated systems, validating the underlying model assumptions on a statistical basis. While this robust agreement reinforces the prospect that large samples of RELHICs could serve as highly compelling probes of the smallest cosmological scales, the predicted scarcity of detectable RELHICs in the local Universe demands careful scrutiny of the robustness of parameter inference for individual systems
If each detection is rare, then gaining an understanding of the following is paramount: the model's object-by-object accuracy, the potential for systematic biases in inferred halo parameters, and the degree to which intrinsic degeneracies---for example, between halo mass and concentration---may limit precision. Furthermore, departures from idealized equilibrium or environmental influences---effects that average out in a large population---are expected to introduce significant scatter or bias in the analysis of individual systems.

The recent discovery of Cloud-9, an isolated ~\ion{H}{I} cloud near M94~\citep{Zhou2023, Benitez-Llambay2023}, underscores the urgency of addressing these outstanding issues. With its substantial~\ion{H}{I} mass and lack of a stellar counterpart down to $M_\star \lesssim 10^{3.5} \, \mathrm{M}_\odot$ (confirmed by deep Hubble Space Telescope WFC/ACS imaging), the object is a prime RELHIC candidate~\citep{Anand2025}. However, its projected proximity to M94 and slightly perturbed morphology raise questions about its dynamical state, potentially violating the core assumptions of hydrostatic equilibrium and the outer pressure that underpin the simple RELHIC model~\citep{Benitez-Llambay2024}.

The complexities presented by Cloud-9 exemplify the very challenges that will be faced by future observations when interpreting the growing population of similar objects. As ongoing surveys with FAST~\citep[e.g.,][]{Zhang2024}, ASKAP~\citep[e.g.,][]{Koribalski2020}, and MeerKAT~\citep[e.g.,][]{deBlok2020,deBlok2024} are poised to discover numerous new candidates, it is imperative to shift the focus of theoretical work from statistical validation to the reliable application of models to individual systems, each with its own unique history and environment.

In this paper, we directly confront this challenge. Using a suite of RELHICs identified in a high-resolution large-volume cosmological simulation, we apply the analytic model of \citetalias{Benitez-Llambay2017} to infer their host halo parameters on an object-by-object basis. We systematically assess the model's performance by testing it against both the spherically averaged 3D gas and~\ion{H}{I} density profiles as well as the intrinsic 2D~\ion{H}{I} column density profiles. By comparing these inferences to the true halo properties known from the simulation, we quantify the bias and determine the fundamental limits of parameter recovery inherent to the method.\footnote{We plan to report on the limitations imposed by observational limitation in a follow-up study.} Thus, we extend previous validation studies from population averages to the single-object regime, establishing the theoretical baseline required for interpreting future observations of systems such as Cloud-9.

This paper is organized as follows. In Sect.~\ref{Sec:Method} we describe our methodology, presenting the analytic model in Sect.~\ref{Subsec:AnalyticModel} and the simulation details in Sect.~\ref{Sec:Simulation}. In Sect.~\ref{Subsec:RelhicSample} we outline the RELHIC selection criteria and sample properties, followed by the construction of the density profiles in Sect.~\ref{Subsec:DensityProfiles}. In Sect.~\ref{Sec:Results} we present the parameter inference results derived from gas density profiles, starting with an illustrative example in Sect.~\ref{Subsec:IllustrativeExample} and extending the analysis to the full sample in Sect.~\ref{Subsec:HaloParametersAll}. We investigate the physical origin of the observed biases in Sect.~\ref{Subsec:Environment} and quantify the impact of the environment in Sect.~\ref{Subsec:EnvImpact}. In Sect.~\ref{Sec:HIprofiles} we assess the performance of the model using realistic \ion{H}{I} distributions and column densities. Finally, we discuss the results in Sect.~\ref{Sec:Discussion} and end by summarizing our conclusions in Sect.~\ref{Sec:Conclusions}.

\section{Method}
\label{Sec:Method}

\subsection{Analytic model}
\label{Subsec:AnalyticModel}

As discussed, RELHICs are structurally simple systems, consisting of dark matter halos filled with largely diffuse gas. Their simplicity stems from the state of this gas: it is partially photoionized and in thermal equilibrium with the external UVB. This equilibrium creates a tight coupling between the gas temperature and density, establishing a well-defined equation of state, $T(\rho)$. This relationship allowed us to construct a relatively simple model that describes the average gas structure of RELHICs.

Following~\citetalias{Benitez-Llambay2017}, we assumed that the gas is spherically symmetric and in hydrostatic equilibrium. This condition is described by the hydrostatic equilibrium equation
\begin{equation}
\frac{1}{\rho} \frac{dP}{dr} = -G\frac{M(<r)}{r^2},
\end{equation}
where $P$ and $\rho$ are the gas pressure and density, respectively, and $M(<r)$ is the total mass enclosed within radius $r$.

Following~\citetalias{Benitez-Llambay2020}, we defined the following dimensionless variables: 
$\tilde{M} = M/M_{200}$, 
$\tilde{T} = T/T_{200}$, 
$\tilde{r} = r/r_{200}$, 
and $\tilde{\rho} = \rho / \bar\rho_{\rm bar}$,
where $M_{200}$ and $r_{200}$ are the dark matter halo virial mass and virial radius, respectively, and $\bar\rho_{\rm bar}$ is the cosmic mean baryon density. 
The virial temperature of the halo is defined as
\begin{equation}
    T_{200} = \frac{\mu m_{\rm p}}{2 k_{\rm B}}\, V_{200}^2,
\end{equation}
where $V_{200} = (G M_{200} / r_{200})^{1/2}$ is the halo virial circular velocity, and $k_{\rm B}$, $m_{\rm p}$, and $\mu$ are, respectively, the Boltzmann constant, the proton mass, and the mean molecular weight, which, for simplicity, we assumed to be $\mu=0.6$. Under this transformation, and assuming $\tilde{T} = \tilde{T}(\tilde{\rho})$, the previous equation reads\begin{equation}
\label{Eq:dimensionless}
\left(\frac{\tilde{T}}{\tilde{\rho}} + \frac{d\tilde{T}}{d\tilde{\rho}}\right) \frac{d\tilde{\rho}}{d\tilde{r}} = -2 \frac{\tilde{M}(<\tilde{r})}{\tilde{r}^2}.
\end{equation} In these systems, the gravitational potential is overwhelmingly dominated by dark matter, so we approximated the total enclosed mass, $\tilde{M}(<\tilde{r})$, by the dark matter halo mass, $\tilde M(<\tilde r) = \tilde{M}_{\rm dm}(<\tilde{r})$. We modeled the dark matter distribution using the standard Navarro--Frenk--White (NFW) density profile \citep{Navarro1996a, Navarro1997}, whose enclosed mass is
\begin{equation}
    \tilde{M}(<\tilde{r}) = \frac{1}{f_c} \left[\ln(1 + c\tilde{r}) - \frac{c\tilde{r}}{1 + c\tilde{r}}\right],
    \label{Eq:MassProfile}
\end{equation}
where $f_c = \ln(1 + c) - c/(1 + c)$, and $c$ is the concentration parameter of the halo.

To obtain the gas density profile $\tilde{\rho}(\tilde{r})$, we integrated Eq.~(\ref{Eq:dimensionless}). To solve this equation, three inputs are required: (1) the halo mass profile, which depends on concentration, (2) the equation of state, $\tilde{T}(\tilde{\rho})$, governing the gas temperature-density relation (we discuss this in Sect.~\ref{Subsec:RELHICs-EoS}), and (3) a boundary condition. Following \citetalias{Benitez-Llambay2017}, we set this boundary condition by equating the gas density at infinity to the mean density of the universe. With these components, the model uniquely determines the gas density profile for each dark matter halo. 

Once the density and temperature profile of RELHICs are known, we derived the total gas mass, $M_{\rm gas}$, the neutral hydrogen density profile, $\rho_{\rm HI}$, the total neutral hydrogen mass, $M_{\rm HI}$, and the neutral hydrogen column density profile, $N_{\rm HI}$. Throughout this work, we assumed a primordial hydrogen mass fraction $X=0.75$. For the neutral hydrogen fraction, $f_{\rm HI}$, we approximated it using the fitting function of~\cite{Rahmati2012}.

\begin{figure*}
    \centering
    \includegraphics[width=0.9\textwidth]{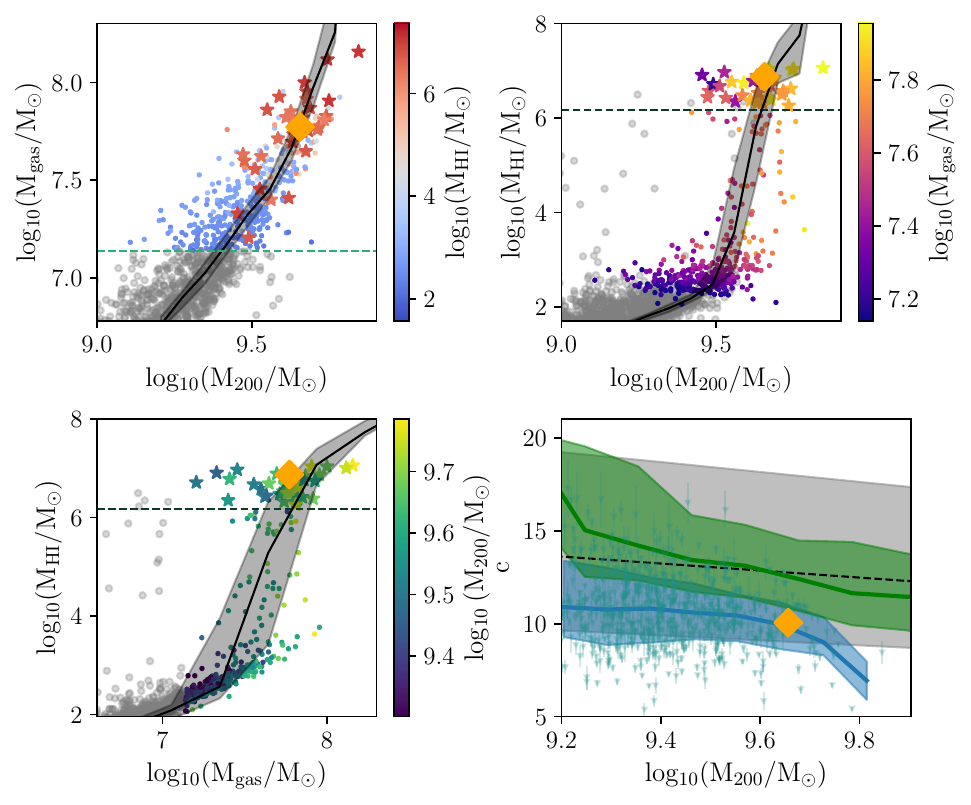}
    \caption{{\it Top left}: Gas mass vs. halo mass for the RELHIC sample. Points are colored by \ion{H}{I} mass. The solid black line shows the expected gas-halo mass relation from the~\citetalias{Benitez-Llambay2017} model using the effective equation of state of the RELHICs and the median RELHIC mass-concentration relation; the shaded region indicates the expected scatter in gas mass from the scatter in RELHICs' concentrations. Stars mark~\ion{H}{I}-rich RELHICs. The large orange diamond symbol indicates an individual RELHIC example, displayed by the thick orange line in Fig.~\ref{fig:AllGasDensiyProfiles}. 
    {\it Top right}: \ion{H}{I} mass vs. halo mass. Points are colored by gas mass. The dotted line shows the threshold~\ion{H}{I} mass adopted to divide the systems between~\ion{H}{I}-rich ($M_{\ion{H}{I}} \gtrsim 10^6\,M_{\odot}$) and~\ion{H}{I}-poor ($M_{\ion{H}{I}} \lesssim 10^6\,M_{\odot}$). The solid black line and shaded region indicates, as before, the expected \ion{H}{I}-halo mass relation from the~\citetalias{Benitez-Llambay2017} model. 
    {\it Bottom left}: \ion{H}{I} mass vs. gas mass, colored by halo virial mass. The solid black line and shaded region mark the~\citetalias{Benitez-Llambay2017} predictions together with the scatter arising from the scatter in concentration. {\it Bottom right}: Halo mass-concentration relation for the selected RELHIC sample (blue) compared to the relation for galaxies (green). Individual RELHICs are shown by the small triangles. The black line indicates the expected trend from the \citet{Ludlow2016} relation, with the shaded regions representing the 16th–84th percentile intervals. The blue solid line and shaded region indicate the median and the scatter of the mass-concentration relation for our RELHIC sample.}
    \label{Fig:SelectedHalos}
\end{figure*}

\subsection{Halo parameters from the density profile of the RELHICs}
\label{Subsec:HaloParameters}

The framework introduced above also allows the dark matter distribution to be constrained from an observed gas density profile. To see this, it is useful to consider the isothermal case, whose analytic solution is
\begin{equation}
\tilde{\rho}(\tilde{r}) = \exp \left\{ \frac{2}{\tilde T} \frac{\ln(1 + c\tilde{r})}{f_c \tilde{r}} \right\}.
\end{equation} This expression shows that as $\tilde{r} \rightarrow 0$, the central density converges to the characteristic value,
\begin{equation}
\tilde{\rho}_0 = \exp\left(\frac{2 c}{\tilde{T}\,f_c}\right).
\end{equation} Consequently, the central density of RELHICs, $\tilde{\rho}_0$, encodes key information about the mass and concentration of their host halo. However, these two structural parameters are degenerate with respect to the central density.

Specifically, $\tilde{\rho}_0$ is an increasing function of both halo mass and concentration.\footnote{$\tilde{\rho}_0$ is an increasing function of $c$ only for $c\gtrsim 2.16$. For $c\lesssim 2.16$ the function is decreasing.} Because of this, a single fixed value of $\tilde{\rho}_0$ does not correspond to a unique halo. Instead, it defines a family of solutions in the mass-concentration plane where a higher halo mass can be compensated by a lower concentration, and vice versa, to produce identical central densities. This degeneracy leads to an intrinsic anticorrelation between the inferred values of halo mass and concentration when a measurement of $\tilde{\rho}_0$ is the primary constraint.
This degeneracy, however, arises from relying on a single central data point. In principle, the degeneracy between $c$ and $M_{200}$ can be broken by utilizing information from the full gas profile, $\tilde{\rho}(\tilde{r})$. The shape of the gas density profile is sensitive to both $M_{200}$ and $c$, and because these parameters influence the profile's shape differently, measuring the gas density at two distinct radii (e.g., at the center $\tilde{r}_1 = 0$ and at an external radius $\tilde{r}_2 = \tilde{r}_{\text{ext}}$) provides two independent constraints. By finding the ($M_{200}$, $c$) pair that simultaneously satisfies both constraints, one can uniquely determine the host halo mass and its concentration.

Although utilizing two discrete radial points is theoretically sufficient to break the $c$-$M_{200}$ degeneracy, this method is highly sensitive to localized fluctuations. Small deviations from a perfect equilibrium profile or artifacts at the chosen radii can introduce significant bias into the derived parameters. A more robust approach would involve leveraging the information encoded across the entire radial profile. 

We therefore employed Bayesian nested sampling~\citep{Skilling2004} to determine the posterior probability distribution for the parameter set ($M_{200}$, $c$) given all available data points. Specifically, we utilized the public Python package {\tt dynesty}~\citep{Speagle2020}, a state-of-the-art implementation of the dynamic nested sampling algorithm~\citep{Higson2019}. Unlike standard Markov chain Monte Carlo methods, dynamic nested sampling calculates the Bayesian evidence by sorting individual samples to estimate the fraction of prior volume they enclose, ensuring the parameter space is explored both efficiently and thoroughly. By applying {\tt dynesty} to the full RELHIC gas profile, we can robustly estimate the complete posterior probability distribution, identifying the global best fit while naturally marginalizing over localized variations.

\subsection{The simulation}
\label{Sec:Simulation}
We used a high-resolution smoothed-particle-hydrodynamics (SPH) cosmological simulation performed with the {\tt P-GADGET3} code~\citep[based on][]{Springel2005}. The simulation follows the evolution of a periodic cubic cosmological volume with a side length of $20$ comoving megaparsecs. Initial conditions were generated at $z=127$ using the {\tt MUSIC} code \citep{Hahn2011}. The run adopted the physics modules of the {\tt EAGLE RECAL} model~\citep{Schaye2015,Crain2015} and assumed a $\Lambda$CDM cosmology consistent with the \citet{Planck2014} parameters.

The simulation was run with $2\times 1024^3$ particles, resulting in a dark matter particle mass of $m_{\rm dm} \approx 2.8 \times 10^5\, \mathrm{M}_\odot$ and an initial gas particle mass of $m_{\rm gas} \approx 5.26 \times 10^4\,\mathrm{M}_\odot$. The Plummer-equivalent gravitational softening, $\epsilon$, never exceeds $\epsilon \approx 195 \rm \ pc$ for the gas particles, and corresponds to approximately 1$\%$ of the mean interparticle separation. 

The simulation includes key baryonic physics relevant for galaxy formation. Gas cooling and heating are implemented element-by-element following the tables of \citet{Wiersma2009}. For the pristine gas in our low-mass halos, cooling is dominated by hydrogen and helium processes. The intergalactic medium is photoionized and heated by the~\citet{Haardt2012} redshift-dependent cosmic UVB, which is activated at the epoch of reionization, $z_{\mathrm{re}}=11.5$.

Star formation is modeled by stochastically converting gas particles into star particles based on the {\tt EAGLE} pressure-based Kennicutt-Schmidt relation. This process is applied only to gas particles that exceed a hydrogen number density threshold of $n_{\rm H, th} = 1.0\,\mathrm{cm^{-3}}$, which is independent of metallicity and ensures that the gas is self-gravitating before forming stars.

Dark matter halos were first identified at $z=0$ using a standard friends-of-friends (FoF) algorithm with a linking length $b=0.2$ \citep{Davis1985}. The {\tt HBT+} code \citep{Han2018} was then used to identify all self-bound (sub)halos within these FoF groups and trace their merger histories through time. This procedure generated a comprehensive catalog for each snapshot, providing positions, velocities, and the gravitationally bound particle lists (gas, dark matter, and stars) for all ``central'' and ``satellite'' halos.

\begin{figure}
    \centering
    \includegraphics[width=1\columnwidth]{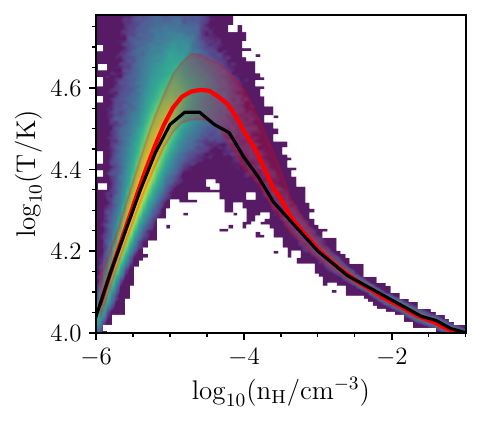}
    \caption{Temperature–density relation for all gas particles belonging to our RELHIC sample. The heatmap illustrates the particle number density on a logarithmic scale, computed using a $100 \times 100$ grid. The bins are equally spaced in log-space, covering the ranges $-6 \le \log_{10}(n_{\text{H}} / \text{cm}^{-3}) \le  0$ and $4 \le \log_{10}(T/ \text{K}) \lesssim 4.8$. The color map represents the number of particles per bin,  ranging from 10 (blue) to a peak density of $\approx 7 \times 10^{3}$ (yellow). The solid red line shows the running median (our effective equation of state), while the shaded red region spans the 16th–84th percentiles. For comparison, the solid black line shows the analytic equation of state reported by~\citetalias{Benitez-Llambay2017}.}
    \label{Fig:EOS}
\end{figure}

\subsection{RELHIC sample}
\label{Subsec:RelhicSample}

Our goal is to estimate the halo parameters of simulated RELHICs at the present day. To this end, we identified 406 RELHICs in our simulation at $z=0$ as starless ``central'' dark matter halos that contain more than 300 bound gas particles but no stars inside the virial radius.

This sample represents almost the entire starless halo population above $\log{(M_{200}/\mathrm{M_\odot})}=9.5$ (see Fig.~\ref{Fig:SelectedHalos}), yielding a number density of $0.050 \pm 0.003\,\mathrm{Mpc^{-3}}$.  Additionally, all the systems are compatible with being in hydrostatic equilibrium, displaying a total gas velocity dispersion of $\rm 2.57^{+2.29}_{-0.96}\, km/s$ (where uncertainties represent the 16th and 84th percentiles of the distribution). Since we used  the same simulation of~\citetalias{Benitez-Llambay2020}, our RELHIC sample forms as discussed by these authors. Specifically, RELHICs are systems that failed to form stars prior to reionization, with their gas being kept in thermal equilibrium with the external UVB and in hydrostatic equilibrium with the dark matter halo.

For comparison, previous studies report $\sim 10$--$30$ starless halos with $M_{\rm gas} \gtrsim 10^{5}\,\mathrm{M_\odot}$ within a spherical volume of radius $\sim 3.5\,\mathrm{Mpc}$, corresponding to a number density of $\sim 0.055\,\mathrm{Mpc^{-3}}$ \citep{Benitez-Llambay2017, Garcia-Bethencourt2026}. By adopting their lower gas mass threshold ($10^{5}\,\mathrm{M_\odot}$) and partitioning our full simulation box into 64 subvolumes of comparable size, we recover a median density of $0.097^{+0.048}_{-0.048}\,\mathrm{Mpc^{-3}}$ (where uncertainties represent the 16th and 84th percentiles). Therefore, our underlying population is comparable to those found in previous works.

The conservative lower limit on gas mass ensures that the gas component is reasonably well resolved, allowing us to use the gas density profile as a tracer of the underlying dark matter potential. The RELHICs analyzed here, therefore, contain a total gas mass, $M_{\rm gas} \gtrsim 1.6 \times 10^{7}\,\mathrm{M_{\odot}}$. However, we caution that imposing a fixed gas mass threshold introduces a selection bias at the low-mass end of our sample. For halos with virial masses $M_{200} \lesssim 10^{9.5}\,\mathrm{M_{\odot}}$, this criterion preferentially selects the tail of gas-rich systems for a given halo mass, excluding the population of typical, and lower-gas-mass halos that fall below our ``detection'' limit.

Figure~\ref{Fig:SelectedHalos} shows the relations between gas mass, halo mass, and \ion{H}{I} mass for the selected RELHICs. In the top-left panel, individual systems are colored according to their~\ion{H}{I} mass, which depends on the density and temperature structure of each system. The mass of~\ion{H}{I} was calculated using the fitting functions of~\citet{Rahmati2012}. The top-right and bottom-left panels show the~\ion{H}{I} mass as a function of halo mass and virial gas mass, colored by their virial gas mass and halo mass, respectively.

This figure demonstrates that RELHICs transition from being \ion{H}{I}-poor to \ion{H}{I}-rich at $M_{\rm 200} \gtrsim 3 \times 10^{9}\,\mathrm{M_{\odot}}$. This sharp transition motivated us to define \ion{H}{I}-rich systems as those with a neutral hydrogen mass of $M_{\rm HI} \gtrsim 10^6\, M_{\odot}/h$. Although this threshold is arbitrary, it allows us to divide the RELHIC sample into two distinct populations, seen in the top-right and bottom-left panels. 

Notably, some of these objects deviate significantly from the mean \ion{H}{I}-gas/halo mass relation predicted by the~\citetalias{Benitez-Llambay2017} model, indicated by the black line in the panels (see Appendix~\ref{App:HI-rich deviations} for further details). The model was evaluated using the effective equation of state of the RELHICs, which we measure directly from the simulation (see Fig.~\ref{Fig:EOS}), and using the concentrations of the RELHICs shown in the bottom-right panel. The shaded region around the models indicates the expected variation in gas mass arising from scatter in the halo mass–concentration relation of the RELHICs.

\subsection{Mass-concentration relation of the RELHICs}
\label{Subsec:mass-cocentration}

We estimated the concentration of each dark matter halo by fitting its cumulative mass profile,\footnote{Fitting the cumulative rather than the actual density profile provides greater numerical stability.} which depends only on the concentration parameter (see Eq.~\ref{Eq:MassProfile}). 
We performed the fit using the Levenberg-Marquardt algorithm implemented in the {\tt scipy} {\tt curve\_fit} function \citep{Virtanen2020}. For each halo, we adopted the best-fitting value as the halo concentration estimate, with the uncertainty corresponding to the square root of the variance of the parameter estimation. We referred to these concentration estimates as the ``ground truth'' for our RELHIC sample. We adopted the virial mass measured by {\tt HBT} as the ground-truth halo mass for each RELHIC.

The bottom-right panel of Fig.~\ref{Fig:SelectedHalos} shows the dark matter halo concentration of our RELHIC sample as a function of halo mass. The blue symbols represent individual RELHICs, while the solid blue line and shaded region indicate the running median and the 16th–84th percentile range, respectively. For comparison, we also show the mass–concentration relation for central galaxies identified in our simulation. Although these central galaxies are not the focus of this work, we include them to highlight that RELHICs are intrinsically biased toward low concentrations compared to the expected average $\Lambda$CDM concentration at a fixed halo mass. 

While the luminous central galaxy population is consistent with the~\citet{Ludlow2016} mass-concentration relation (see black dashed line), RELHICs lie systematically below this relation. This offset is expected, as RELHICs, particularly the most massive ones, inhabit halos that assembled significantly later than the halos of similar present-day mass that host luminous galaxies. Halo concentration is strongly correlated with formation time, as a halo's characteristic density (and thus its scale radius, $r_{\rm s}$) reflects the mean density of the Universe at its assembly epoch~\citep[e.g.,][and references therein]{Wechsler2002, Ludlow2016}. Consequently, late-forming halos of RELHICs naturally have larger scale radii and, therefore, lower concentrations at a fixed mass~\citep[see, e.g.,][for a similar result]{Garcia-Bethencourt2026}. 

\subsection{Equation of state of the RELHICs}
\label{Subsec:RELHICs-EoS}

Gas accreting onto RELHICs remains relatively cold, as their low virial temperatures ($T_{200} < 10^4\,\mathrm{K}$) preclude shocks from heating the gas to temperatures exceeding $10^4\,\mathrm{K}$. This gas is instead photoheated by the external UVB, and its resulting thermal state depends on density, $n_{\text{H}}$. For $n_{\text{H}} \gtrsim 10^{-4.5} \text{ cm}^{-3}$, thermal timescales are short compared to the Hubble time ($t_{\text{th}} \ll t_{\text{Hubble}}$), allowing the gas to reach an equilibrium temperature where UVB photoheating balances radiative cooling. At lower densities, $t_{\text{th}} > t_{\text{Hubble}}$, and thermal equilibrium is not achieved. Instead, the gas temperature is established by the competing effects of photoheating and adiabatic cooling due to the expansion of the Universe~\citep[e.g.,][]{Haehnelt1996, Theuns1998}. 

However, the lack of thermal equilibrium at low densities does not preclude pressure equilibrium.  The sound-crossing horizon, $l_{\text{sc}} \approx c_{\rm s} / H_0$, is $\approx 300 \text{ kpc}$ (assuming a typical sound speed $c_{\rm s} \approx 20 \text{ km s}^{-1}$ and a Hubble constant $H_0 = 67.7 \text{ km s}^{-1} \text{ Mpc}^{-1}$). Since the virial radii of RELHICs ($R_{200} < 40 \text{ kpc}$) are much smaller than $l_{\rm sc}$, the sound-crossing time across the halos is $t_{\text{sc}} \ll t_{\text{Hubble}}$. The gas within RELHICs can therefore be safely assumed to be in pressure equilibrium.

The arguments above imply that the gas in RELHICs follows a well-defined, effective equation of state. While \citetalias{Benitez-Llambay2017} showed that this equation of state can be derived from these simple arguments, we took a more pragmatic approach here and measured the effective equation of state directly from the simulation. To this end, we stacked the density and temperature of all gas particles belonging to RELHICs and defined the $T(\rho)$ relation as the median of the resulting temperature distribution for each density. 

Figure~\ref{Fig:EOS} shows this temperature–density relation for all gas particles belonging to our RELHICs sample. The color map indicates the particle number density, ranging from low (blue) to high (yellow) values. The red solid line tracks the running median, which we adopted as the ``effective'' equation of state used throughout this work, while the shaded region spans the 16th–84th percentiles of the distribution. For comparison, the black solid line shows the analytic RELHIC equation of state reported by~\citetalias{Benitez-Llambay2017} (see their Table A1). The difference between our empirically derived equation of state and that of~\citetalias{Benitez-Llambay2017} is minimal, driven mostly by the interpolation~\citetalias{Benitez-Llambay2017} used to connect the low- and high-density regimes of their analytic estimates.

Although Fig.~\ref{Fig:EOS} reveals some scatter around the median effective equation of state, the temperature variations at a fixed density are relatively small. Consequently, adopting the median equation of state provides a sensible approximation for our parameter inference pipeline.

\subsection{Density profiles of the RELHICs}
\label{Subsec:DensityProfiles}

We constructed the gas density profiles for our RELHIC sample using SPH particle densities from the simulation snapshots. To derive the neutral hydrogen content, we assumed a primordial hydrogen mass fraction of $X = 0.75$ and calculated the~\ion{H}{I} density via the~\cite{Rahmati2012} fitting function, which accounts for self-shielding against the UVB. The profiles were computed in 27 logarithmically spaced spherical shells within the range $-2.0 \le \log_{10}(r/R_{200}) \le 1.3$.\footnote{Virial radii are taken from {\tt HBT}. For RELHICs, $R_{200} \lesssim 40 \rm \  kpc$.} For each shell, we defined the density as the median of each bin. We defined the innermost radius as the minimum radius that contains at least 10 particles, and we excluded bins with a radius smaller than this; this was done to mitigate the effects of the limited resolution of our simulation, which can bias parameter inference (we assess the impact of resolution in Appendix~\ref{App:Resolution}). 

Figure~\ref{fig:AllGasDensiyProfiles} displays the resulting profiles, colored by virial mass. As shown by the oblique dashed line, the innermost radius we can robustly measure varies systematically across the sample, scaling approximately as $r_{\rm min} \propto n_{\rm H}^{-1/3}$. This scaling is largely the direct consequence of the finite resolution of our simulation. For a fixed gas particle mass, the only way to sample lower densities is by increasing the spatial distance between the SPH gas particles, inherently limiting our ability to robustly measure the density in the innermost regions. Consequently, while the inner regions of gas-rich halos are well resolved down to $\approx 0.3$ kpc, we can only probe regions beyond $5-7$ kpc in gas-poor systems. Figure~\ref{fig:AllGasDensiyProfiles}, therefore, suggests that this resolution constraint will affect the recovery of dark matter halo parameters for low-mass systems, particularly the halo concentration, which is highly sensitive to the central density. Indeed, the median concentration of RELHICs in our simulation is $c \approx 10$, implying the need to resolve the gas density profile on scales $r_{\rm min} \lesssim r_{200} / c \ll 4$ kpc.

\begin{figure}
    \centering
    \includegraphics[width=1\columnwidth]{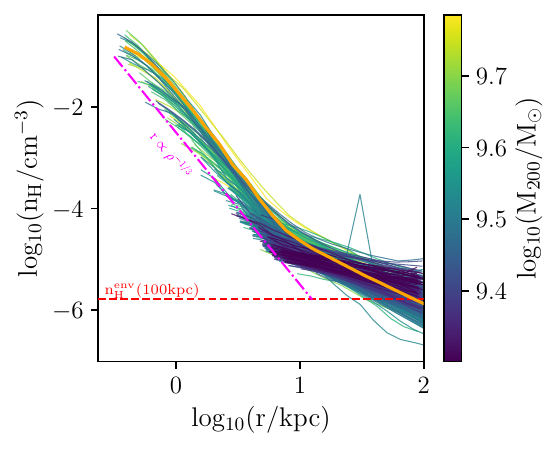}
    \caption{Gas density profiles for our RELHIC sample, colored by halo virial mass. The thick orange line shows the profile of the RELHIC example highlighted in Fig.~\ref{Fig:SelectedHalos} (orange diamond) and discussed in Sect.~\ref{Subsec:IllustrativeExample}. The horizontal red dashed line indicates the expected density at 100 kpc from the~\citetalias{Benitez-Llambay2017} model for a halo similar to the example.}
    \label{fig:AllGasDensiyProfiles}
\end{figure}

Our analysis of neutral hydrogen was limited to the 35 systems classified as~\ion{H}{I}-rich ($M_{\rm HI} \gtrsim 10^{6} \text{ M}_{\odot}/h$), as identified in Fig.~\ref{Fig:SelectedHalos}. We calculated the~\ion{H}{I} density profile for each system using the same spherical shell procedure described above for the total gas.

We computed the intrinsic~\ion{H}{I} column density maps using the {\tt py-sphviewer} code~\citep{sphviewer}, which projects the mass distribution onto a 2D regular Cartesian grid using the SPH approximation. In particular, we considered a grid with a resolution of $\approx 170$ pc per pixel centered on each RELHIC. We extracted profiles from this grid using 27 circular annuli centered at the radius $R$, logarithmically spaced over the range $-2.0 \le {\rm log}_{10} (R/ {R_{200}}) \le 1.5$. Consistent with our volume density analysis, the column density and uncertainty in each annulus were defined as the median and the standard deviation, respectively.

\section{Results}
\label{Sec:Results}

Having established the methodology for defining our RELHIC sample and measuring its intrinsic properties, in this section we focus on inferring host halo properties from the gas distributions of RELHICs. We begin by discussing an illustrative example that highlights our methodology, after which we apply the same framework to the entire RELHIC sample in Sect.~\ref{Subsec:HaloParametersAll}.

\subsection{Illustrative example: Halo parameters from the gas distribution}
\label{Subsec:IllustrativeExample} 

We begin with an assessment of the performance of our methodology for retrieving dark matter halo mass and concentration solely from the gas distribution. We focus on the example RELHIC highlighted by the large orange diamond symbol in Fig.~\ref{Fig:SelectedHalos}, which inhabits a dark matter halo of mass $M_{200} \approx 4.5 \times 10^{9}\,{\rm M_{\odot}}$ and concentration $c=10.05 \pm 0.41$. Its gas density profile is shown by the thick line in Fig.~\ref{fig:AllGasDensiyProfiles}. For this system, we employed {\tt dynesty}, using only the gas density profile as the input constraint. 

The \texttt{dynesty} algorithm determines the best-fitting halo mass and concentration based on the RELHIC model introduced in Sect.~\ref{Subsec:AnalyticModel}. We assumed a Gaussian log-likelihood function defined as
\begin{equation}
    \ln \mathcal{L} = -\frac{1}{2} \displaystyle\sum_{i}^{N} \left[ \frac{(y_{{\rm obs},i} - y_{i})^2}{s^2} + \ln(2\pi s^2) \right] ,
\end{equation}
where $y_{{\rm obs},i}$ is the logarithmic density of the RELHIC measured in the simulation, $y_{i}$ is the logarithmic density of the RELHIC model, both evaluated at radius $r_{i}$, $N$ is the number of bins in which we evaluate the profile, and $s$ is treated as a free nuisance parameter. This approach is necessary because the exact uncertainties associated with the SPH density profile are difficult to quantify on an object-by-object basis. Since nested sampling is highly sensitive to the adopted uncertainties, treating $s$ as a free parameter allows the sampler to naturally capture the intrinsic scatter of the data, enabling us to safely marginalize over it. Unless stated otherwise, we adopted the same prior for $s$ throughout this work.

We assigned a log-uniform prior for $s$ in the range $-3 \le \log_{10} s \le 3$. For the physical parameters, we adopted a log-uniform prior for the virial mass, $5 \times 10^8 \le M_{200} / \rm M_{\odot} \le 10^{10}$, and a uniform prior for the concentration, $2 \le c \le 30$. Finally, we defined the optimal parameter values as the medians of the resulting marginalized posterior distributions, estimating their uncertainties via the 16th and 84th percentiles.

\begin{figure}[]
\centering
\includegraphics[width=1\linewidth]{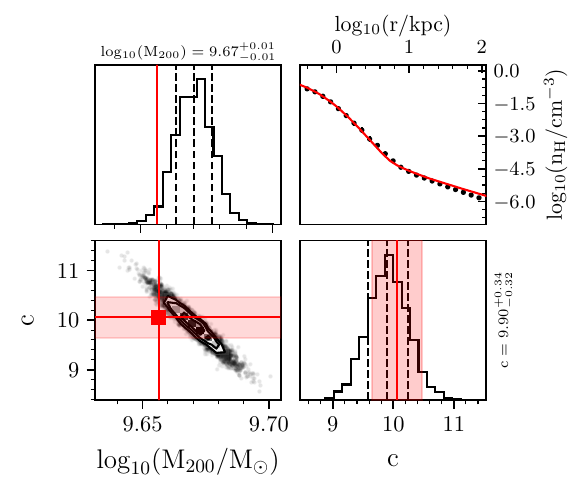}
\caption{Posterior distributions of mass and concentration of a RELHIC example, obtained by applying nested sampling to its gas density profile. The red lines and shaded regions mark the ground-truth values from the simulation. The top-right panel compares the input gas density profile (black symbols) with the best-fitting RELHIC model (solid line). The recovered nuisance parameter $s$ is $\approx 0.1$.}
\label{fig:Posteriorior_gas}
\end{figure}

Figure~\ref{fig:Posteriorior_gas} presents the posterior distributions for the halo virial mass and concentration. The diagonal panels display the marginalized one-dimensional distributions. In the bottom-left panel, which shows the joint posterior, the intersection of the red lines marks the ``ground-truth'' values derived from the simulation, with the shaded bands representing their associated uncertainties.

As anticipated in previous sections, we observed a strong anticorrelation between virial mass and concentration. This degeneracy stems largely from the constant central gas density characteristic of RELHICs, which is determined by a combination of both parameters. Despite this intrinsic degeneracy, the joint analysis of the posterior distributions constrains the halo parameters remarkably well. The inferred values for the mass of $M_{200} = (4.7 \pm 0.1) \times 10^{9} \rm \ M_{\odot}$ and concentration, $c=9.9^{+0.34}_{-0.32}$, are in good agreement with the ground truth. The mass is consistent with the ground truth within $2\sigma$, whereas the concentration is spot-on. For reference, the recovered nuisance parameter is $s\approx 0.1$.

Finally, the top-right panel shows the gas density profile of this system (black dots) together with the best-fitting model (red solid line). Although a small systematic discrepancy in the recovered profile is observed at large distances, indicating a systematic departure from the~\citetalias{Benitez-Llambay2017} model outside the halo, the overall agreement is excellent. This example thus validates our methodology for inferring host halo properties from the gas density profile. We next apply this framework to the entire RELHIC sample.

\subsection{Halo parameters from the gas distribution for the entire RELHIC sample}
\label{Subsec:HaloParametersAll}

Having tested our method on a single system, we then extended the analysis to the entire RELHIC sample. For each object, we computed the gas density profile following the procedure outlined in Sect.~\ref{Subsec:DensityProfiles}. These profiles were then processed with {\tt dynesty}, adopting the same likelihood and priors used in the illustrative example (log-uniform for mass, uniform for concentrations). We defined the inferred mass and concentration as the medians of the posterior distributions, with uncertainties corresponding to the 16th and 84th percentiles.

\begin{figure}
    \centering
    \includegraphics[width=1\linewidth]{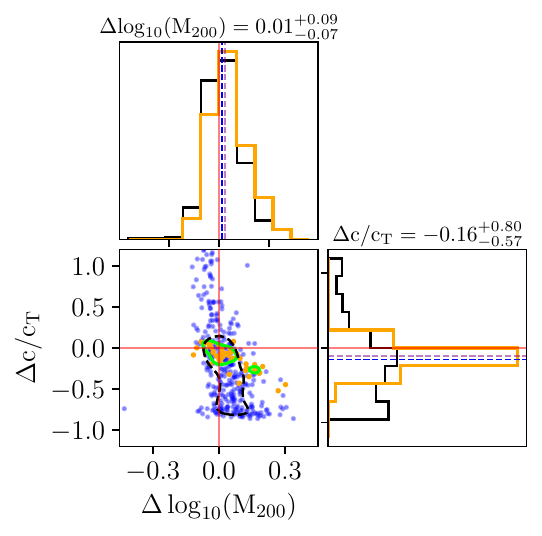}
    \caption{Distribution of the recovered halo parameters relative to the ground truth. For the virial mass, we define the logarithmic deviation as $\Delta \log_{10}(M_{200}) = \log_{10}(M_{200}/M_{200,\rm T})$, while for the concentration we show the fractional error, $\Delta c/c_{\rm T} = c/c_{\rm T} - 1$. The top and right panels show the marginalized one-dimensional histograms, whereas the central panel displays the joint two-dimensional distribution. The blue points and black histograms represent the full RELHIC sample, while the orange points and histograms highlight the subset of ``well-resolved'' systems, i.e., those RELHICs for which the inner density profile can be probed below 1 kpc. Red lines indicate the unbiased reference values.} 
    \label{fig:GasHist}
\end{figure}

Figure~\ref{fig:GasHist} illustrates the accuracy of our halo parameter recovery for the full RELHIC sample. We show the relative error in concentration, $\Delta c / c_{\rm T} = c/c_{\rm T}-1$, against the logarithmic error in virial mass $\Delta {\rm log}_{10} (M_{200}) = {\rm log}_{10}(M_{200}/M_{200,\rm T})$, where the T subscripts denote the ground-truth values derived from the simulation. The top and right panels display the corresponding marginalized distributions. To explicitly test the impact of the finite numerical resolution discussed in Sect.~\ref{Subsec:DensityProfiles}, we separated ``well-resolved'' systems (where the density profile can be probed below 1 kpc) from the rest.
\begin{table}[h] 
    \caption{Accuracy of the recovered parameters from the total gas profiles using nested sampling for the full RELHIC sample and for the ``well-resolved'' ($r_{\rm min} < 1$ kpc) subsample.}
    \centering
    \addtolength{\tabcolsep}{5pt}
    \begin{tabular}{lcc}
    \toprule
    Parameter & Full sample & $r_{\rm min} < 1$ kpc \\
    \midrule
    $\Delta \log_{10}(M_{200})$ & $0.01^{+0.09}_{-0.07}$  & $0.03^{+0.13}_{-0.07}$ \\
    \addlinespace[0.5em] 
    $\Delta c/c_{\rm T}$         & $0.16^{+0.80}_{-0.57}$  & $0.11^{+0.17}_{-0.11}$ \\
    \addlinespace[0.5em]
    $\log_{10}(s)$              & $-1.31^{+0.31}_{-0.33}$ & $-1.00^{+0.22}_{-0.21}$ \\
    \bottomrule
    \end{tabular}
    \label{tab:AccuracyDensityFix}
\end{table}

This figure demonstrates that the recovery of the halo mass is remarkably robust and largely unbiased across the ensemble, irrespective of the simulation's ability to track the profile all the way to the very central regions. Indeed, the median of the marginalized distribution for $\Delta \log_{10}(M_{200})$ is consistent with zero for both the whole sample and the ``well-resolved'' subsample, with an intrinsic scatter of $\approx 25 \%$.

In contrast, the recovered concentrations exhibit a strong dependence on resolution. When evaluating the full sample, the median of the recovered concentrations exhibits a slight bias, underestimating the true values by $\approx 20\%$. However, isolating the ``well-resolved'' systems (indicated by the orange subset) significantly decreases both this systematic bias and the associated scatter by approximately a factor of 1.5 , and 4, respectively. This confirms that the inability to resolve the innermost density of low-mass systems forces our framework to infer a lower concentration. Consequently, we interpret this bias as partly driven by the simulation's finite resolution rather than a fundamental limitation of the analytic model. We discuss this further in Appendix~\ref{App:Resolution}, where we explore the impact of the choice of the innermost radius of the gas density profile on the recovery of halo mass and concentration. In Table~\ref {tab:AccuracyDensityFix} we summarize the accuracy and precision in the recovery of dark matter halo parameters for the whole sample and the subsample of ``well-resolved'' RELHICs, together with the recovered value for the nuisance parameter, $s$.

While Fig.~\ref{fig:GasHist} and Appendix~\ref{App:Resolution} indicate that resolution accounts, to some extent, for the bias in concentration, a nonnegligible scatter remains on a per-object basis for both mass ($\approx 25 \%$) and concentration ($\approx 50 \%$) even among ``well-resolved'' systems. We have verified that these deviations do not stem from poor fits to the gas density profiles, but rather from the inability of the underlying~\citetalias{Benitez-Llambay2017} model to perfectly reproduce the simulated gas distribution when strictly fixed to the true halo parameters, particularly in the outer parts (as already hinted by the illustrative example discussed previously).

\begin{figure}
    \centering
    \includegraphics[width=\columnwidth]{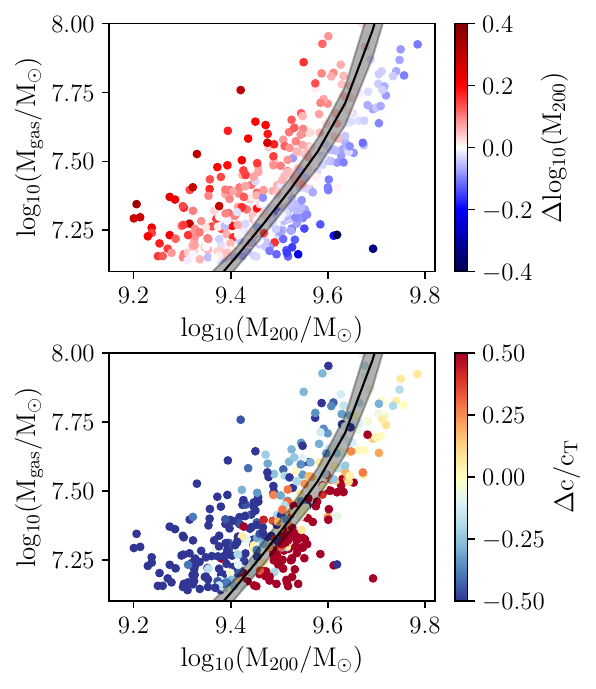}
    \caption{Virial gas mass vs. true dark matter halo mass for our RELHIC sample, colored by the relative accuracy in recovered mass (top) and concentration (bottom). The solid black line indicates the~\citetalias{Benitez-Llambay2017} gas vs. halo mass relation, with the shaded region representing the scatter in gas mass expected from the intrinsic dispersion in the concentration parameter of our simulated RELHICs, as measured in the bottom-right panel of Fig.~\ref{Fig:SelectedHalos}.}
    \label{Fig:Figure6}
\end{figure}

To investigate this issue further, we examined how the recovery accuracy depends on the intrinsic properties of individual systems. Figure~\ref{Fig:Figure6} displays the true gas mass versus the true halo mass of RELHICs, colored by the relative error in the inferred halo mass (top) and concentration (bottom). At a fixed halo mass, we find that RELHICs with overestimated halo masses contain more gas than the~\citetalias{Benitez-Llambay2017} model predicts (indicated by the solid line). Conversely, systems where the halo mass is underestimated are gas-deficient relative to the model. The concentration exhibits an inverse trend: At fixed halo mass, gas-rich systems yield systematically lower-than-expected concentrations, while gas-poor systems yield higher concentrations.

These strong correlations suggest that the scatter in the recovered halo parameters is not stochastic but rather driven by intrinsic variations in the gas content relative to the idealized model. Since the~\citetalias{Benitez-Llambay2017} model produces a specific gas-to-halo mass relation, any ``excess'' or ``deficit'' gas in a simulated RELHIC is naturally interpreted by the fitting procedure as evidence of a deeper (or shallower) gravitational potential. This leads to a systematic overestimation (or underestimation) of the halo mass to provide the necessary force to balance the inferred pressure. 

This intrinsic scatter also explains the asymmetric bias observable at the low-mass end of our sample. By selecting RELHICs above a fixed gas mass threshold, we introduce a selection effect for halos with $M_{200} \lesssim 10^{9.5} \rm \ M_{\odot}$. At these masses, we preferentially include the gas-rich tail of the distribution, missing the gas-poor systems that fall below our resolution limit. Because these selected low-mass systems are systematically gas-rich relative to the model median, our fitting procedure systematically overestimates their halo masses. Due to the inherent anticorrelation between mass and concentration in the profile fitting (as seen in Sect. \ref{Subsec:IllustrativeExample}), this mass overestimation forces a corresponding underestimation of the concentration, explaining the dominance of red (blue) points at the low-mass end of the top (bottom) panel of Fig.~\ref{Fig:Figure6}.
We explore the physical origin of these intrinsic, per-object variations in our simulation next.

\subsection{The origin of the scatter: Environmental effects}
\label{Subsec:Environment}

\begin{figure*}[t]
    \centering
    \includegraphics[width=0.9\linewidth]{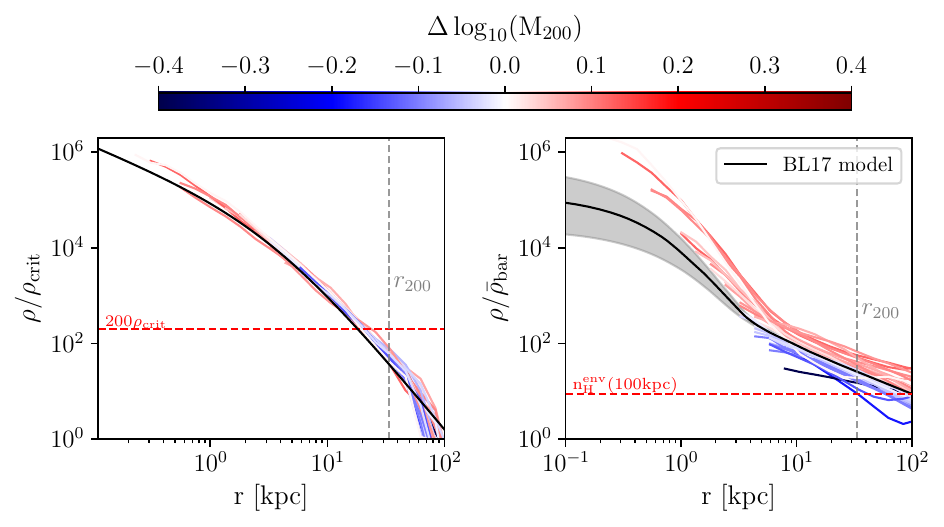}
    \caption{Density profiles of bound dark matter particles (left) and gas (right) for RELHICs within the narrow range of virial mass, $9.58 < \log(M_{200}/ \rm M_{\odot}) < 9.62$, colored by the logarithmic error in the recovered halo mass. Solid black lines indicate the halo and gas density for a RELHIC of mass and concentration equal to the median values of the subsample, following the~\citetalias{Benitez-Llambay2017} model. The shaded regions indicate the 16th and 84th percentiles. The horizontal dashed red lines indicate an overdensity of 200 in the left panel and the expected density at 100~kpc as predicted by the~\citetalias{Benitez-Llambay2017} model in the right panel. The vertical lines mark the virial radius of the median halo. Note that the dark matter density profiles are well described by an NFW profile.}
    \label{fig:Environment}
\end{figure*}

Figure~\ref{Fig:Figure6} suggests that the accuracy of the recovered halo parameters is tightly coupled to the total gas content of the system. Specifically, at a fixed true halo mass, RELHICs that are gas-rich relative to the~\citetalias{Benitez-Llambay2017} model prediction yield overestimated halo masses, while gas-poor systems yield underestimated masses. To understand the physical origin of this diversity, we now focus on determining why halos of identical mass can host such disparate gas reservoirs.

To address this issue, we isolated a subsample of RELHICs within a narrow range of virial mass, $9.58 \lesssim {\rm log}_{10} (M_{200} / {\rm M_{\odot}} ) \lesssim 9.62$, effectively eliminating halo mass as a variable. Figure~\ref{fig:Environment} presents the spherically averaged density profiles of the bound dark matter (left panel) and gas (right panel) particles to these systems.\footnote{While unbound dark matter substructures can pass through the virial outskirts of these systems, they do not meaningfully impact the inferred halo parameters because they are typically less massive than RELHICs and do not contain gas that can change the external pressure. Because RELHICs are spherical and in virial equilibrium, the gravitational potential governing their gas distribution is largely established by their enclosed inner mass.} The curves are colored according to the logarithmic error in the recovered virial mass, $\Delta \log_{10} (M_{200})$.

The left panel reveals that the dark matter profiles of these systems are remarkably similar. There is no systematic difference in the inner structure (or concentration) of the halos that correlates with the recovered mass error. This result confirms that the scatter in gas content is not driven by subtle variations in the true gravitational potentials of the individual dark matter halos.

In contrast, the gas density profiles in the right panel exhibit large diversity. Systems for which the halo mass is overestimated (red lines) display gas densities systematically higher than the sample average at all radii, while systems with underestimated masses (blue lines) are systematically underdense. The solid black line indicates the theoretical profile predicted by the~\citetalias{Benitez-Llambay2017} model for a halo with the median mass and concentration of this subsample. The model prediction lies perfectly in the middle of the distribution, confirming that it accurately captures the behavior of an average RELHIC.

A clue to the origin of this diversity lies in the asymptotic behavior of the gas profiles at large radii ($r \gtrsim 100$ kpc). The ``red'' profiles in the right panel of Fig.~\ref{fig:Environment} do not converge to the expected~\citetalias{Benitez-Llambay2017} density (indicated by the red horizontal dashed line); instead, they flatten at significantly higher values. This suggests that these specific systems reside in locally overdense environments compared to the ones expected by the \citetalias{Benitez-Llambay2017} model. Conversely, the ``blue'' profiles drop below the model prediction in the outskirts, signaling an underdense local environment. 

\begin{figure}
    \centering
    \includegraphics[width=1\linewidth]{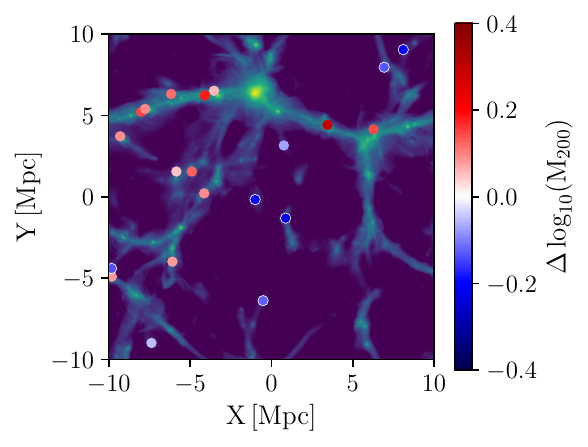}
    \caption{Slice of width $\approx 700\,\mathrm{kpc}\,$ from our large-volume cosmological simulation showing the projected gas density distributions with the positions of the RELHICs overlaid. Individual systems are colored according to their logarithmic deviation in the recovered halo mass. Systems for which the halo mass is overestimated inhabit preferentially denser regions close to larger galaxies and filaments. Conversely, systems for which the halo mass is underestimated reside in lower-density regions, preferentially far away from massive galaxies or filaments.}
    \label{fig:Slice}
\end{figure}

To visually verify this hypothesis, Fig.~\ref{fig:Slice} presents a projected slice of the gas density field from our simulation, overlaid with the positions of the RELHICs. Systems are colored by the logarithmic deviation in the recovered halo mass, following the same scheme as previous figures. A clear spatial segregation is evident: RELHICs for which the halo mass is overestimated (red symbols) are preferentially located in or near high-density structures, indicated by the bright yellow regions. Conversely, systems with underestimated halo masses (blue symbols) are typically found in underdense regions, shown in dark blue. This visualization suggests that the local environment modulates the gas properties of individual RELHICs. A similar environmental dependence is also reported by \citet[see their Fig. 6]{Lee2024}. They find that starless halos residing $\sim 300-500 \, \mathrm{kpc}$ from the nearest filaments experience an environment approximately $10$ times denser than the mean baryonic density of the Universe, whereas those located much further away ($\sim 30-50 \, \mathrm{Mpc}$) inhabit environments with densities roughly $0.1$ times the cosmic mean.

\begin{figure}
    \centering
    \includegraphics[width=\linewidth]{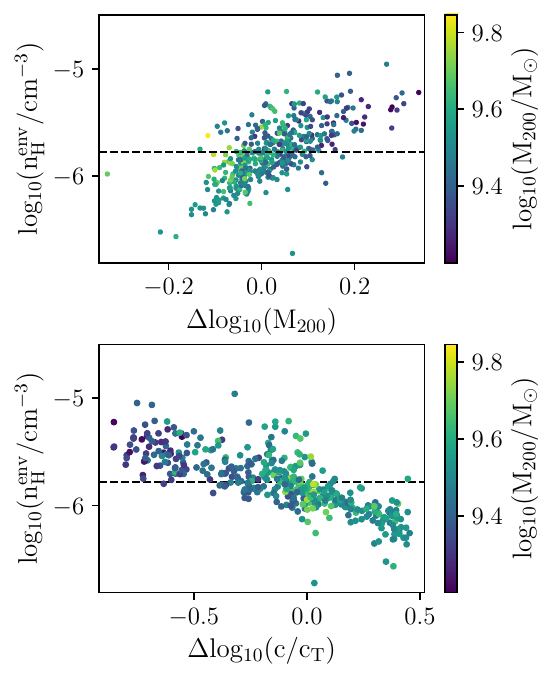}
    \caption{Correlation between environment density and parameter recovery accuracy for the full RELHIC sample. The y-axis shows the gas density measured at $r=100$ kpc, serving as a proxy for the local environment. The horizontal dashed line marks the expected~\citetalias{Benitez-Llambay2017} density at that location, given a boundary condition such that the density is equal to the mean baryon density at infinity. The top panel shows that the environment density correlates positively with the logarithmic error in recovered halo mass. The bottom panel shows that the environment density correlates negatively with the concentration error. Points are colored by the true halo mass (indicated by the color bar). These figures suggest that the bias in individual objects is, in part, driven by the mismatch between the actual environmental density and the cosmic mean assumed by the model.}
\label{Fig:FigEnvironmentDensityAll}
\end{figure}

To generalize this result to the full RELHIC sample, we quantified the local environment for every object in Fig.~\ref{Fig:FigEnvironmentDensityAll}. We measured the gas density at a radial distance of $r \approx 100$ kpc (a scale larger than twice the virial radius of the most massive RELHICs) and plotted it against the recovery accuracy in halo mass (top) and concentration (bottom). The horizontal dashed line indicates the expected density given the boundary condition assumed by the \citetalias{Benitez-Llambay2017} model. Quantitatively, at $r \approx 100 \, \mathrm{kpc}$, we found a median gas density of $\rm log(n_{H}^{env}/cm^3)=-5.79^{+0.28}_{-0.24}$. For comparison, \citet{Garcia-Bethencourt2026} defines the environmental density differently, measuring the gas density within the spherical shell $R_{200} < r < 7 R_{200}$, and reports a median value of the environments of the RELHICs of $\rm log(n_{H}^{env}/cm^3)=-5.91^{+0.32}_{-0.45}$. When we adopted their exact definition for our sample, we recovered an even more compatible median density of $\rm log(n_{H}^{env}/cm^3)=-5.90^{+0.35}_{-0.31}$. The excellent agreement between these values not only confirms the environmental similarities between the two samples but also demonstrates that both methods for estimating the local environmental density yield virtually identical results.

We observed a striking correlation that aligns with our hypothesis. Systems residing in overdense environments consistently yielded positive mass residuals (overestimated $M_{200}$) and negative concentration residuals. Conversely, systems in underdense environments yielded underestimated masses and overestimated concentrations. Crucially, the intersection of the data cloud with the zero-bias axis occurred close to the~\citetalias{Benitez-Llambay2017} density, albeit with large scatter. This suggests that (i) the~\citetalias{Benitez-Llambay2017} model is accurate when its boundary assumptions are met and (ii) that part of the scatter in the recovered host dark matter parameters is driven by the diversity of environmental densities sampled by RELHICs in our simulation ~\citep[see, e.g.,][for a similar result]{Zheng2025}.

\subsection{The impact of environment on halo parameter estimation}
\label{Subsec:EnvImpact}

The results described above suggest that the local environment modulates the gas content of RELHICs, which in turn affects the inference of the dark matter halo parameters. This occurs because the boundary condition directly influences the hydrostatic equilibrium solution. A RELHIC embedded in a dense environment experiences a higher external pressure, which compresses the gas and raises the density throughout the entire system, thereby increasing the total gas mass. 

Our inference method, based on the~\citetalias{Benitez-Llambay2017} model, assumes a fixed boundary condition corresponding to the cosmic mean baryon density, $\bar\rho_{\rm bar}$. When presented with a gas profile ``elevated'' by environmental pressure, the model interprets the excess density as evidence of a deeper potential well. Consequently, it converges on a halo mass larger than the true value to generate the extra gravitational force necessary to explain the higher gas density. The inverse applies to systems in underdense environments.

To quantify this effect, we performed a controlled experiment. We generated synthetic gas density profiles for a fixed dark matter halo of virial mass, $M_{200} = 10^{9.6}\ {\rm M}_\odot$, and concentration, $c=10$, embedded in regions with different background densities spanning two orders of magnitude relative to the cosmic mean,\footnote{These values are comparable to the scatter observed in gas density in the outer parts of RELHICs in Fig.~\ref{fig:AllGasDensiyProfiles}.} i.e., in the range $0.1 \le \rho_{\rm env}/\bar \rho_{\rm bar} \le 10$. We then applied our {\tt dynesty} inference pipeline to these synthetic profiles to recover the halo parameters.

We illustrate the individual best-fitting models for representative underdense, average, and overdense environments in Appendix~\ref{App:EnvironmentFits} (see Fig.~\ref{fig:Environment_theory_3profiles}). As expected, when the environment matches the cosmic mean, {\tt dynesty} recovers the true halo parameters perfectly. In nonaverage environments, the model---forced to assume a boundary at $\bar\rho_{\rm bar}$---struggles to capture the asymptotic behavior at large radii. Notably, however, all best-fitting profiles match the data almost perfectly inside the halo ($r \lesssim R_{200} \approx 30$ kpc), regardless of the accuracy of the final parameter estimation. 

\begin{figure}
    \centering
    \includegraphics[width=1\linewidth]{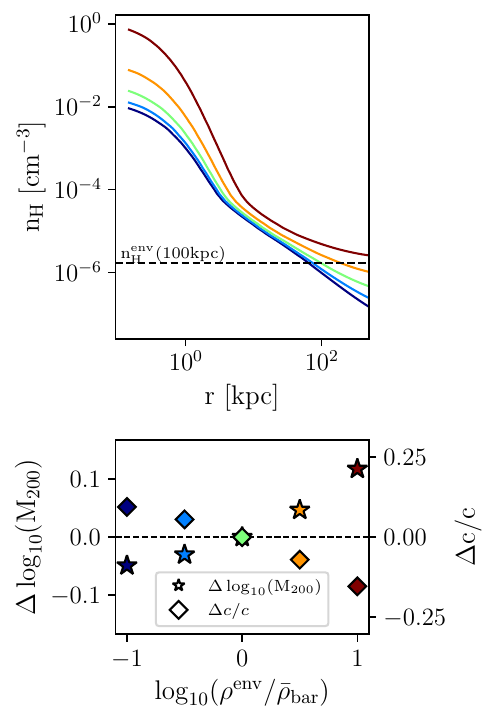}
    
    \caption{Effects of environmental density on the inference of halo parameters under the ~\citetalias{Benitez-Llambay2017} model. The top panel shows synthetic gas density profiles for a fixed RELHIC of virial mass $M_{200} = 10^{9.6} \ \rm M_{\odot}$ and concentration, $c=10$, embedded in environments of different density. Profiles are colored according to the background overdensity relative to the mean baryon density (see the color bar). The bottom panel shows the relative error in the derived halo mass (left axis) and concentration (right axis), as a function of environment overdensity, resulting from applying our {\tt dynesty} pipeline together with the~\citetalias{Benitez-Llambay2017} model.}
    \label{fig:Environment_theory}
\end{figure}

Figure~\ref{fig:Environment_theory} summarizes the results of this experiment together with examples inhabiting intermediate density environments. The top panel compares the synthetic profiles of all these models, demonstrating that a denser environment not only shifts the outer normalization, but also alters the shape of the profile, particularly in the outskirts, making it difficult, in principle, to disentangle halo parameters from environment. 

The bottom panel of this figure explicitly quantifies the bias in the parameter recovery for these systems. We found a clear, monotonic trend: the logarithmic error in the recovered virial mass, $\Delta \log_{10}( M_{200})$ (star symbols; scale on the left), correlates positively with the environment density contrast. Conversely, the concentration parameter (diamond symbols; scale on the right) exhibits a corresponding anticorrelation; as the environmental density increases, our method systematically underestimates the concentration. This behavior arises from the intrinsic degeneracy between mass and concentration in the RELHIC profile fitting; {\tt dynesty} increases the mass to match the ``elevated'' gas density but must simultaneously lower the concentration to prevent the central density from exceeding the observed values. 

Since the scatter within the explored values is comparable to the scatter in the recovered values of the ``well-resolved'' systems of our sample for both variables (see Table~\ref{tab:AccuracyDensityFix}), we can safely conclude that the scatter in the $M_{\rm gas}$-$M_{200}$ relation is largely driven by the environmental modulation of the boundary pressure. While the assumption of a universal cosmic background is statistically valid for the population average, the local environment of individual systems introduces a systematic floor (or ceiling) to the gas density that the simple analytic model mimics by adjusting the halo mass and its concentration. 

\section{Halo parameters from~\ion{H}{I} distributions}
\label{Sec:HIprofiles}

The analysis in the previous sections focused on the total gas density profiles. However, total gas is not directly observable; in practice, radio observations trace the neutral hydrogen component, which in turn traces the innermost structure of RELHICs. We now investigate whether the results obtained from the analysis of the total gas distribution hold when using~\ion{H}{I} as the sole tracer of the gravitational potential.

We focus exclusively on the subsample of 35~\ion{H}{I}-rich RELHICs ($M_{\rm \ion{H}{I}} \gtrsim 10^6 \rm \ M_{\odot}/h$) defined in Sect.~\ref{Subsec:RelhicSample}. These are the systems most likely to be detected by current and future radio surveys. 
We begin by analyzing their spherically averaged 3D~\ion{H}{i} density profiles to establish a theoretical baseline before moving to more realistic projected column density profiles in the following section.

\subsection{Inference from 3D~\ion{H}{I} density profiles}
\label{Subsec:HIDensity}

\begin{figure}
    \centering
    \includegraphics[width=\linewidth]{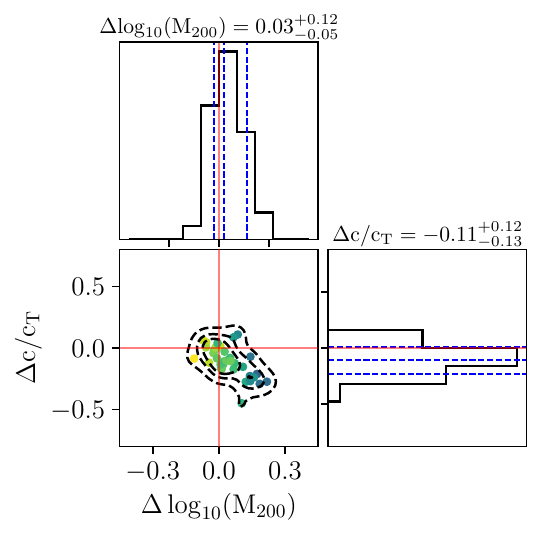}
    \caption{Distribution of the recovered halo parameters relative to the ground truth, inferred from the 3D spherically averaged~\ion{H}{I} density profiles of the~\ion{H}{i}-rich RELHIC subsample ($M_{\rm HI} \gtrsim 10^6 \text{ M}_\odot/h$). The top and right panels show the marginalized one-dimensional distributions for the logarithmic error in virial mass ($\Delta \log_{10} (M_{200})$) and the fractional error in concentration ($\Delta c/c_{\rm T}$), while the central panel displays their joint two-dimensional distribution. Red lines indicate the unbiased reference values. The~\ion{H}{I} density provides a highly accurate tracer of the potential, yielding a median mass recovery offset of just $0.03$ dex. The concentration is slightly biased (median offset of $-0.11$).}
    \label{fig:nHIHist_test}
\end{figure}

Following the methodology in Sect.~\ref{Subsec:DensityProfiles}, we computed the spherically averaged~\ion{H}{I} density profiles for the~\ion{H}{I}-rich RELHIC subsample. We then applied our {\tt dynesty} inference pipeline using the~\citetalias{Benitez-Llambay2017} model, accounting for the ionization state of the gas. We adopted the same priors for the halo mass, concentration, and for the nuisance parameter $s$ described previously.

Figure~\ref{fig:nHIHist_test} presents the distribution of the recovered halo parameters relative to the ground truth. This figure is the~\ion{H}{I}-equivalent of Fig.~\ref{fig:GasHist}, but it is limited to the~\ion{H}{I}-rich systems. This figure confirms that~\ion{H}{I} is an effective tracer of the underlying dark matter potential, yielding constraints comparable to, if not tighter than, those obtained from the total gas profiles. The mass recovery is largely unbiased, with a median offset of just $\Delta {\rm log}_{10} (M_{200}) = 0.03$ dex, while the concentration is only slightly underestimated by $\approx 11 \%$. The systematics in the concentration recovery are largely driven by the few low-mass (but otherwise~\ion{H}{I}-rich) systems of the sample, which correspond to the darker data points in the central panel. These are systems that contain an unusually high fraction of~\ion{H}{I} for their gas mass compared to the majority of halos of similar mass, and also in light of the~\citetalias{Benitez-Llambay2017} model (see Fig.~\ref{Fig:SelectedHalos}). As suggested by the experiments we carried out in Appendix~\ref{App:Resolution}, it is likely that the concentration recovery for these systems is highly affected by the limited resolution of our simulation. The fact that the mass and concentration for the most massive~\ion{H}{I}-rich systems (indicated by the lighter data points) are largely unbiased supports this interpretation.

As established in Sects. \ref{Subsec:Environment} and \ref{Subsec:EnvImpact}, some of the scatter in the derived halo parameters is likely driven by the modulations of the local density surrounding these systems. Because our \ion{H}{I}-rich selection inherently requires a high gas density to cross the self-shielding threshold, it also introduces a selection effect. To appear in this subsample, lower-mass halos must reside in overdense environments that compress their gas, leading the model to slightly overestimate their mass. Conversely, the most massive systems in our simulation happen to reside in slightly underdense environments, leading to minor mass underestimations.

Nevertheless, despite these environment modulations, the~\ion{H}{I} profiles successfully constrained both the halo mass and concentration to well within a factor of 2 of the ground truth on a per-object basis. Having established this robust baseline performance on 3D profiles and identified the physical drivers of the scatter, we now turn to the analysis of projected (observationally more realistic) 2D profiles.

\subsection{Inference from 2D \texorpdfstring{\ion{H}{I}}{HI} column density profiles}
\label{Subsec:HIColumnDensity}

Having quantified the performance of the model in 3D, we now take a step closer to observations by considering the projected gas distribution. We computed the intrinsic~\ion{H}{I} column density profiles, $N_{\rm HI}(R)$, for our subsample of~\ion{H}{I}-rich systems by integrating the 3D distribution along the line of sight, as detailed in Sect.~\ref{Subsec:DensityProfiles}. We then repeated the {\tt dynesty} inference, this time fitting the projected analytic model to these 2D profiles.

\begin{figure}
    \centering
    \includegraphics[width=\linewidth]{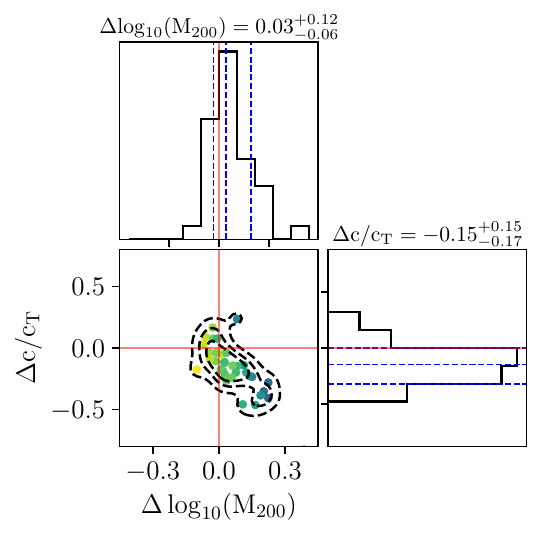}
    \caption{Same as Fig.~\ref{fig:nHIHist_test} but for halo parameters inferred from the 2D projected intrinsic~\ion{H}{I} column density profiles of the \ion{H}{I}-rich RELHIC subsample. While the virial mass remains robustly constrained (median offset of 0.03 dex), line-of-sight projection effects smooth out the detailed radial profile structure, causing our inference to systematically underestimate the concentration parameter for the ensemble (median offset of $-0.15$).}
    \label{fig:NcolumnHIHist_test}
\end{figure}

Figure \ref{fig:NcolumnHIHist_test} presents the accuracy of the parameter recovery from these projected profiles. The format is identical to the previous figure, displaying the joint and marginalized distributions for the relative errors in virial mass and concentration across the~\ion{H}{I}-rich subsample.

Comparison with the 3D results (Fig. \ref{fig:nHIHist_test}) reveals two important facts. Firstly, the per-object biases are robust to projection. The marginalized mass distribution remains skewed toward slight overestimation (a median $\Delta \log_{10} (M_{200})$ of $0.03$ dex). Secondly, projection exacerbates the parameter degeneracy. While the median bias in the recovered halo mass remains small ( $0.03$ dex) , the recovery of the halo concentration degrades more. The median concentration bias drops from $\Delta c/c_{\rm T} \approx -0.11$ in the 3D analysis to $\approx -0.15$ in the projected analysis. However, as was the case for the 3D profiles, this systematic bias is largely driven by the few low-mass RELHICs that make it to our sample (darker symbols in the central panel).

This expected shift likely occurs because the column density at any projected radius $R$ includes contributions from gas at all physical radii $r \ge R$. This line-of-sight integration smooths out the detailed radial features of the density profile that are more critical for distinguishing between the effects of halo mass and concentration. In 3D, a massive, low-concentration halo and a less massive, high-concentration halo might produce gas profiles that differ subtly in shape. However, in projection, these differences are largely washed out. Consequently, the fitting procedure relies primarily on the overall normalization of the projected profile to infer the mass, forcing the concentration to adjust to match the central column density value.

Ultimately, these results demonstrate that while line-of-sight projection masks the structural details needed for accurate and precise concentration estimates, intrinsic~\ion{H}{I} column density profiles remain a robust theoretical baseline for observational probes. We stress that these are idealized projections; introducing realistic observational limitations---such as instrumental noise, spatial beam smearing, and finite spectral resolution---will inevitably add further complexity. Even so, within this intrinsic and idealized framework, the~\citetalias{Benitez-Llambay2017} model successfully constrained the true underlying halo mass excellently to within $\approx 0.1$ dex despite the environmental diversity inherent in a cosmological volume. 

\section{Discussion}
\label{Sec:Discussion}

Our analysis suggests that the recovery of dark matter halo parameters from RELHIC gas profiles is fundamentally modulated by the diversity in gas content of RELHICs, which in turn depends on the local environment. The hydrostatic equilibrium of the gas is maintained not only by the gravitational potential of the halo but also by the boundary pressure exerted by the intergalactic medium. The analytic model of \citetalias{Benitez-Llambay2017} assumes a boundary condition tied to the mean baryon density of the Universe. While statistically valid for a large cosmological population, this assumption inevitably breaks down for individual objects residing in local over- or under-densities.

We found that the ambient density plays a nonnegligible role in parameter estimation. In overdense regions, the elevated external pressure compresses the gas, boosting the central density of RELHICs. When fitting such a profile with a model that assumes a mean-density boundary, the inference pipeline is forced to increase the halo mass to generate the same inner gas profile. Due to the intrinsic $M_{200}-c$ degeneracy, this mass overestimation is compensated for by a lower concentration. This mechanism naturally explains the anticorrelation between recovered mass and concentration residuals observed in Fig.~\ref{Fig:FigEnvironmentDensityAll}.

This environmental dependence has implications for future observational surveys. Firstly, our results revealed a stark selection bias for the lowest mass halos ($M_{200} \lesssim 10^{9.5} M_\odot$). In our simulation, these halos only form sufficient~\ion{H}{I} to cross our detectability threshold ($M_{\rm HI} \geq 10^6 M_\odot/h$) if they reside in significantly overdense environments. Consequently, the population of low-mass RELHICs that may be accessible to observers will not be a representative random sample but rather a biased subset of environmentally compressed systems. 

Secondly, if these future detections are modeled using standard hydrostatic assumptions (i.e., neglecting local environmental pressure), their dark matter masses will be systematically overestimated. This is particularly relevant for recent candidates such as Cloud-9 \citep{Zhou2023, Benitez-Llambay2023}. Given its projected proximity to the massive galaxy M94, it is highly plausible that Cloud-9 experiences an environmental pressure significantly higher than that of a mean environment. Our results suggest that this compression could lead to an inflated inferred mass, potentially masking a smaller underlying dark matter halo. The extremely low concentration derived for Cloud-9 by \citet{Benitez-Llambay2023} may well be a direct signal of this unmodeled boundary effect, but we stress that the evidence of ongoing ram-pressure stripping for Cloud-9 complicates the interpretation~\citep{Benitez-Llambay2024}.
To unlock the full potential of RELHICs as cosmological probes, this environmental degeneracy must be compellingly addressed. We propose that instead of fixing the boundary density to the cosmic mean, inference models should treat the external environmental density ($\rho_{\rm env}$) as a free parameter, using a prior consistent with simulation predictions.
\begin{figure}[h]
    \centering
    \includegraphics[width=\linewidth]{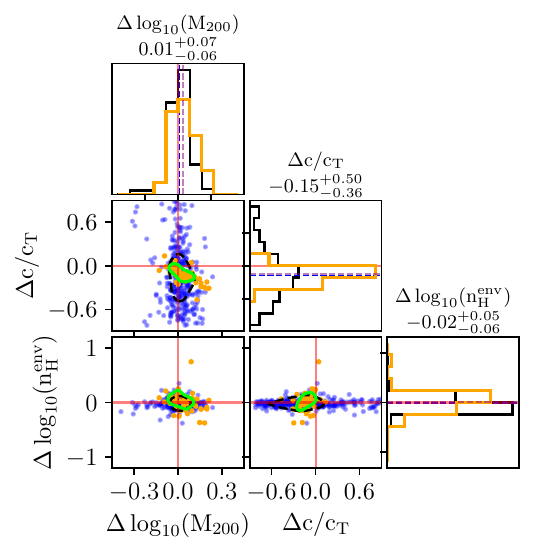}
    \caption{Distribution of the recovered parameters relative to the ground truth from a three-parameter fit to the 3D total gas density profiles of the full RELHIC sample. In this inference, the local environmental density ($n_{\rm H}^{\rm env}$) is treated as a free parameter alongside the halo virial mass and concentration. The top and diagonal panels show the marginalized one-dimensional histograms for the logarithmic error in mass ($\Delta \log_{10}(M_{200})$), the fractional error in concentration ($\Delta c/c_{\rm T}$), and the logarithmic error in the environmental density ($\Delta \log_{10} (n_{\rm H}^{\rm env})$), while the central panels display their joint two-dimensional distributions. The inference adopts identical priors for mass and concentration as in previous sections combined with a log-normal prior for the environmental density centered on the cosmic mean with a standard deviation of 0.5 dex. Red lines indicate the unbiased reference values. Allowing the boundary condition to vary successfully eliminates the systematic mass bias and accurately recovers the true local environment of the systems. As in Fig.~\ref{fig:GasHist}, the blue points and black histograms represent the full RELHIC sample, while the orange points and histograms highlight the subset of well-resolved systems where the inner density profile can be probed below 1 kpc.}
    \label{fig:Figure13}
\end{figure}

To test the viability of this strategy, we performed a final experiment. Returning to the fundamental 3D total gas density profiles of our full sample, we repeated our {\tt dynesty} inference, but this time we allowed the environmental density to vary freely alongside halo mass and concentration. The priors for halo mass and concentration remained identical to those used in our previous inferences. For the environment density, we adopted a log-normal prior distribution, centered at the cosmic mean baryon density, ${\bar\rho}_{\rm bar}$, with a standard deviation of 0.5 dex, which reproduces the scatter observed in Fig.~\ref{Fig:FigEnvironmentDensityAll}.

Figure \ref{fig:Figure13} presents the results of this three-parameter inference. Interestingly, freeing the boundary condition improved the recovered halo mass, reducing the median offset to a negligible $\Delta \log_{10} (M_{200}) = 0.01$ dex. Furthermore, the pipeline successfully recovered the true local environmental density of the systems ($\Delta \log_{10}( n_{\rm H}^{\rm env}) \approx 0.02$) with a very small scatter of only $\approx 0.05$ dex.

Crucially, while a slight systematic underestimation in the concentration persists ($\Delta c/c_{\rm T} \approx -0.15$), the overall scatter in the recovered concentration was noticeably reduced compared to the fixed-boundary case, indicating a substantial gain in the precision of the method. By allowing the boundary condition to move freely, our inference pipeline correctly attributed the behavior of the outer gas to the environment, rather than forcing nonphysical adjustments onto the dark matter halo to compensate. 

Restricting the sample solely to the "well-resolved" systems yielded only a marginal improvement in the accuracy of the already excellently derived virial masses ($\Delta\log_{10}(M_{200}) = 0.04_{-0.07}^{+0.09}$). However, the concentration exhibited a more substantial improvement, reaching $\Delta c/c_{\rm T} = -0.13_{-0.11}^{+0.10}$. While a residual bias persists in the recovered concentration, the precision improved by a factor of 2 compared to inference with a fixed boundary condition. We speculate that the systematic bias in concentration recovery is partly driven by the limited resolution of our simulations and not necessarily due to inherent discrepancies with the~\citetalias{Benitez-Llambay2017} model. We explore this further in Appendix~\ref{App:Resolution}.

In Appendix~\ref{App:DecouplingDensity}, and particularly in Fig.~\ref{Fig:app_rho_free}, we show that by allowing our inference pipeline to recover the background density, the correlation between the outer density of RELHICs and the residuals of the inference of the halo parameters disappeared, further supporting the conclusion that fixing the ambient density artificially ties the inferred halo properties to their local environment. As illustrated in the top panel, treating the background density as a free parameter successfully removed the environmental trend for the halo mass, with the residuals ($\Delta\log_{10} (M_{200})$) cleanly centered around zero across the probed range of $n_{\rm H}^{\rm env}$. However, while decoupling the background density successfully mitigates the environmental dependence, it does not yield entirely unbiased estimates for the concentration. The bottom panel demonstrates that although the systematic tilt with environmental density was gone, an overall offset in the concentration residuals ($\Delta\log_{10}(c/c_{\rm T})$) remained. As argued above, we speculate that this lingering bias is limited by the spatial resolution of the simulations, which may prevent our pipeline from perfectly resolving the innermost halo structure even when the outer boundary conditions are accurately modeled (see Appendix~\ref{App:Resolution}).

The previous exercise indicates that accounting for external pressure is an important component of the inference process. Rather than treating the environmental density as a free parameter, imposing informed priors---or fixing it based on informed observations---breaks degeneracies and improves precision. For specific targets such as Cloud-9, where the local environment can be partially constrained by its proximity to a massive host galaxy, this approach would tighten the structural halo constraints even further.\footnote{We plan to report on the application of our methodology to Cloud-9, taking into account observational complications, in a forthcoming contribution.}

Consequently, when realistic boundary conditions are appropriately incorporated, our results suggest that the~\citetalias{Benitez-Llambay2017} framework proves to be a robust and reliable tool for weighing the dark matter halos of RELHICs. However, properly modeling these environmental effects will be essential for utilizing RELHICs as pristine cosmological probes in the near future. By demonstrating that the true dark matter halo mass can be accurately and precisely recovered when the environmental boundary pressure is physically constrained, we establish a reliable pathway to weigh these systems, thereby enabling robust constraints on mass scales below those of galaxies.

\section{Conclusions}
\label{Sec:Conclusions}

In this work, we performed an object-by-object assessment of the~\citetalias{Benitez-Llambay2017} analytic model for inferring the dark matter halo properties of simulated RELHICs from their gas distributions. By applying the model to a sample of starless halos identified in a high-resolution cosmological hydrodynamical simulation, we quantified the intrinsic accuracy and precision of the underlying dark matter halo parameter recovery and isolated the physical drivers of systematic biases. Our main findings are summarized as follows:

\begin{itemize}
    \item The~\citetalias{Benitez-Llambay2017} model successfully describes the structure of simulated RELHICs on a population-average basis. When applied to the total 3D gas density profiles of the full simulated ensemble, the assumption of hydrostatic equilibrium in a UVB-dominated thermal background provides a robust, largely unbiased theoretical baseline. Crucially, we found that the virial mass is exceptionally well constrained by the gas distribution, recovering the true underlying halo mass with a typical object-by-object scatter of only $\approx 0.1$ dex ($\approx 25\%$).
    
    \item While the population average is largely unbiased, individual systems exhibit systematic residuals that are tightly coupled to their local environment. Our standard inference model assumes a universal boundary condition where the gas density equals the cosmic mean baryon density at large distances. However, RELHICs in locally overdense (underdense) regions contain more (less) gas than the model predicts due to external intergalactic pressure. Failing to account for this localized ambient density leads the inference pipeline to artificially overestimate (underestimate) the halo mass, and correspondingly underestimate (overestimate) the concentration, for environmentally compressed systems.
    
    \item Neutral hydrogen is a highly effective tracer of the dark matter potential, but observational detectability thresholds introduce a selection bias. For the~\ion{H}{I}-rich population ($M_{\rm HI} \gtrsim 10^6 \, \mathrm{M}_\odot/h$) at the lowest mass scales ($M_{200} \lesssim 10^{9.5}\, \mathrm{M}_\odot$), our RELHIC sample was dominated by halos that only formed sufficient~\ion{H}{I} because they were environmentally compressed.
    
    \item Moving from 3D spherically averaged densities to 2D intrinsic column density profiles inevitably degrades the constraining power of the data. Nevertheless, the virial mass remained remarkably robust to projection effects, yielding a median bias of just $\approx 0.03$ dex and maintaining a scatter of $\approx 0.12$ dex. 

    \item The systematic biases introduced by the local intergalactic medium can be effectively mitigated. By treating the local environmental density ($\rho_{\rm env}$) as a free parameter in the nested sampling pipeline, rather than fixing it to the cosmic mean, we successfully recovered the true ambient density of the systems. This approach almost entirely eliminated the systematic bias in the derived halo mass ($\Delta \log_{10} M_{200} \approx 0.01$ dex) and further tightened the typical scatter to $\approx 0.07$ dex, demonstrating that the virial mass can be precisely determined when boundary conditions are appropriately modeled. 
\end{itemize}

These findings have direct implications for the interpretation of~\ion{H}{I}-rich dark systems discovered by ongoing and future radio surveys. Because observational samples will likely be biased toward environmentally compressed systems at the low-mass end, applying the standard~\citetalias{Benitez-Llambay2017} hydrostatic model with fixed cosmic boundaries can systematically inflate their inferred dark matter masses while lowering their concentrations. This is particularly relevant for recent candidates such as Cloud-9, whose projected proximity to M94 implies a nonnegligible local overdensity. For such objects, an unmodeled external pressure could mask a significantly smaller underlying dark matter halo. To properly weigh these systems, inference models must treat the environmental boundary density as a free parameter optimally supplemented by informed observational priors mapping the local cosmic environment.

Although our pipeline increased precision and mitigated the correlation between the inferred halo mass and the environment, a persistent systematic underestimation remains in the recovered halo concentrations. As discussed, our analysis suggests this issue is largely driven by the finite spatial resolution of our cosmological volume, which prevents us from fully resolving the innermost density profiles of low-mass systems. However, follow-up studies are required to definitively assess whether this concentration bias is exclusively driven by resolution or if it stems from intrinsic second-order physical deviations from the~\citetalias{Benitez-Llambay2017} model. This can be addressed by performing targeted very high resolution zoom-in simulations of individual RELHICs identified in our cosmological box, thereby avoiding the computationally prohibitive cost of simulating a full cosmological volume at that resolution.

Finally, we emphasize that the results presented in this work establish a fundamental theoretical baseline. We demonstrated the absolute limits of parameter recovery using idealized noise-free gas distributions extracted directly from simulations. In reality, the observational recovery of these parameters will be further complicated by instrumental limitations. Having isolated the underlying physical and environmental drivers of the inference pipeline here, we plan to report on the performance of this framework when applied to realistic mock observations incorporating, for example, instrumental noise, finite spectral resolution, and beam smearing across a variety of specific instrumental setups in a forthcoming contribution. Ultimately, only through careful exploration of these realistic observational effects, guided by the theoretical baseline established here, will we be able to transform future detections of RELHICs into genuine cosmological probes on yet unexplored mass scales.

\begin{acknowledgements}
ABL acknowledges support by the Italian Ministry for Universities (MUR) program “Dipartimenti di Eccellenza 2023-2027” within the Centro Bicocca di Cosmologia Quantitativa (BiCoQ), and support by UNIMIB’s Fondo Di Ateneo Quota
Competitiva (project 2024-ATEQC-0050). We acknowledge the anonymous referee for constructive comments that improved the presentation of our results. 
\end{acknowledgements}

\bibliographystyle{aa}
\bibliography{bibliography}

\begin{appendix}

\section{Deviations from the BL17 model in HI-rich systems}
\label{App:HI-rich deviations}
In Sect.~\ref{Subsec:RelhicSample}, especially in the top right panel of Fig.~\ref{Fig:SelectedHalos}, we show that the most \ion{H}{I}-rich systems ($M_{\rm HI} \gtrsim 10^{6} \, \mathrm{M_{\odot}}/h$) appear to deviate from the \citetalias{Benitez-Llambay2017} model. The distribution of the total gas velocity dispersion of this subsample is $\rm 2.89^{+2.37}_{-0.97}\, km/s$, where uncertainties represent the
16th and 84th percentiles; therefore, they are compatible with the hydrostatic equilibrium assumption.

\begin{figure}[h]
    \centering
    \includegraphics[width=1\linewidth]{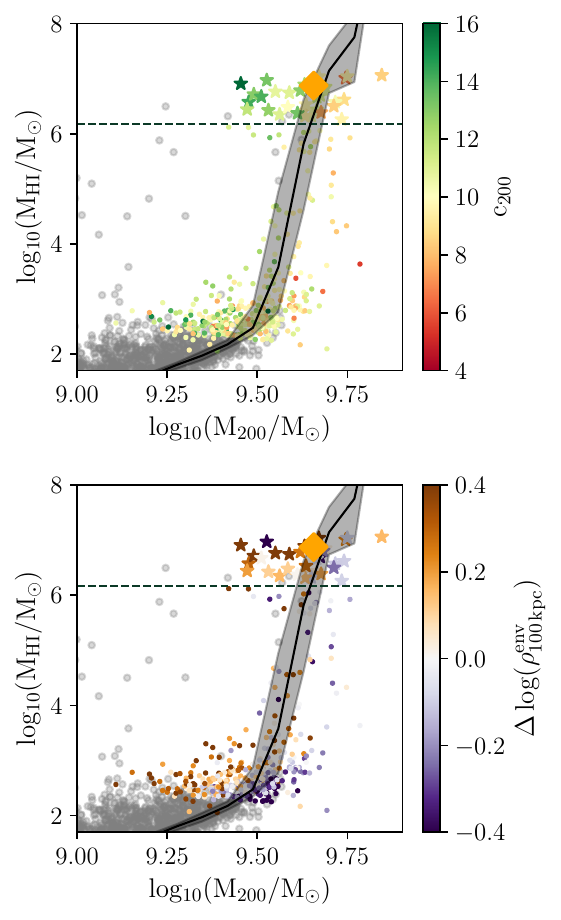}
    \caption{\ion{H}{I}-gas mass relation for the same selected sample of RELHICs of Fig.~\ref{Fig:SelectedHalos}. The dashed horizontal line is the $\rm M_{HI}$ threshold used in this work to define \ion{H}{I}-rich systems, the orange diamond is the example RELHIC analyzed in Sect.~\ref{Subsec:IllustrativeExample}, while the black solid line represents the median \ion{H}{I} mass expected by the \citetalias{Benitez-Llambay2017}  model given the median concentration for each mass bin; the shaded region is the expected scatter due to the concentrations values. {\it Top:} The color code represents the host halo concentration. {\it Bottom:} The color code represents the difference between the density expected by the \citetalias{Benitez-Llambay2017} model and the one measured in the simulation.}
    \label{fig:HI-deviation}
\end{figure}

As shown in Fig.~\ref{fig:HI-deviation}, at lower masses, the systems that depart most significantly from the \citetalias{Benitez-Llambay2017} relation are characterized by denser local environments (bottom panel) and higher concentrations compared to the median ($\rm c_{200} \sim 10$; top panel). This is a natural consequence of their structure: highly concentrated halos possess deeper central potential wells than their lower-concentration counterparts of the same mass. This deeper potential allows them to retain more gas in the center, significantly enhancing their overall neutral \ion{H}{I} content.

\begin{figure}[h]
    \centering
    \includegraphics[width=1\linewidth]{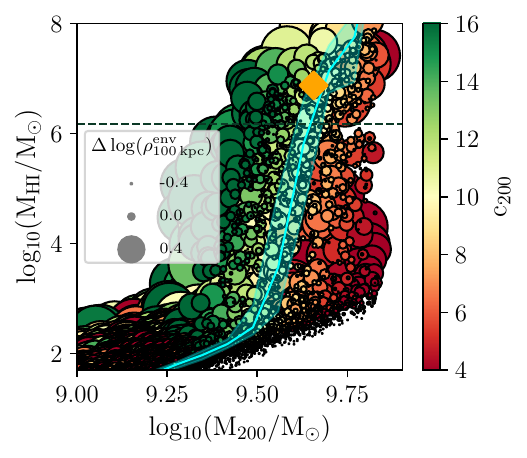}
    \caption{\ion{H}{I} mass vs.\ $M_{200}$ for a generated sample of 5000 mock profiles. The color of each marker represents the halo concentration parameter, while the marker size scales with the local environmental density at 100~kpc. Larger symbols indicate denser local environments. Profiles with central densities exceeding $n_{\rm H} = 5\,\mathrm{cm^{-3}}$ are omitted to represent only the starless halos. For comparison, the dashed horizontal line is the $\rm M_{HI}$ threshold used in this work to define \ion{H}{I}-rich systems. The orange diamond marks the example RELHIC analyzed in Sect.~\ref{Subsec:IllustrativeExample}. Finally, the cyan solid line represents the median \ion{H}{I} mass expected by the \citetalias{Benitez-Llambay2017} model, given the median concentration of the simulated systems in each mass bin; the shaded region shows the expected scatter due to the concentration values.}
    \label{fig:HI-deviation-th}
\end{figure}

To better understand the role of the environment and determine which of the two factors primarily drives the \ion{H}{I} content, we generated 5000 mock gas profiles for halos in the $10^{9}$--$10^{10}\,\mathrm{M_\odot}$ virial mass range, imposing the model boundary condition at 100 kpc. For each profile, the concentration parameter was randomly drawn from a gaussian distribution centered at $c_{200}=10$ with a standard deviation of 5. Simultaneously, the environmental density deviation was drawn from a log-normal distribution centered at 0 with a dispersion of 0.35 dex, comparable to the scatter observed in the simulation data. This deviation was applied to a baseline environmental density of $\rm n_{H}=10^{-6}\,cm^{-3}$, which roughly corresponds to the environmental density expected by the \citetalias{Benitez-Llambay2017} model at 100 kpc from the center. Finally, we discarded any profile where the central gas density exceeds $\rm n_{H}=5\,cm^{-3}$. We chose this upper limit to account for the subgrid star formation criterion, which stochastically converts gas particles into stars following a pressure-dependent Kennicutt-Schmidt relation above a threshold of $n_{\rm H}=1\,\mathrm{cm^{-3}}$. Setting our cut slightly higher allowed us to properly model halos that contain dense star-forming gas that due to stochasticity has not yet formed any stellar particles in the simulation.

It is worth noting that the survival of these highly concentrated, \ion{H}{I}-rich systems as purely starless halos is deeply sensitive to the adopted subgrid star formation threshold. For instance, recent studies analyzing the HESTIA simulations \citep[e.g.,][]{Garcia-Bethencourt2026} do not report similar high-\ion{H}{I} outliers deviating from the median \citetalias{Benitez-Llambay2017} relation. We attribute this apparent discrepancy to the significantly lower star formation density threshold of HESTIA ($n_{\rm H} = 0.13\,\mathrm{cm^{-3}}$) compared to the EAGLE model ($n_{\rm H} = 1.0\,\mathrm{cm^{-3}}$). In models with lower thresholds, the enhanced central gas densities reached by these concentrated halos in dense environments would rapidly trigger star formation. Consequently, these objects would inevitably be converted into dwarf galaxies, naturally removing them from the starless RELHIC sample. Interestingly, in the NIVARIA-LG simulations also analyzed by \cite{Garcia-Bethencourt2026}, for which the star formation density threshold is $10 \rm \ cm^{-3}$, two of these objects are readily visible, reinforcing our interpretation of the role of this subgrid parameter in determining the prominence of this rare population. 

As highlighted in Fig.~\ref{fig:HI-deviation-th}, the mock RELHICs show a clear dependence on halo concentration. Highly concentrated systems ($c_{200} \sim 14$--$16$) form substantial neutral hydrogen at much lower halo masses compared to median-concentration halos ($c_{200} \sim 10$). In contrast, low-concentration systems ($c_{200} \sim 4$--$6$) struggle to form \ion{H}{I} even at higher masses ($\sim 10^{9.75}\,\mathrm{M_\odot}$), where most halos typically start forming stars (see the top-left panel of Fig.~\ref{Fig:SelectedHalos}). They can only form \ion{H}{I} if they live in dense environments ($\Delta \log(\rho_{100\,\mathrm{kpc}}^{\rm env}) \gtrsim 0.4$). Furthermore, if we look at halos with the same concentration and \ion{H}{I} mass, the lower-mass halos are found in denser environments, while the higher-mass halos live in underdense regions. This happens because a dense environment increases the total amount of gas in the entire halo. Concentration, instead, mainly deepens the central potential well and compresses the gas in the core. In short, concentration and environment work together to build up the \ion{H}{I} mass. The environment controls the total amount of gas available to the halo. However, at a fixed halo mass, concentration is the main driver. This is because concentration directly boosts the gas density in the central region, which is exactly where the neutral hydrogen resides.

We can now use these insights to explain the apparent discrepancies in our simulation. The halos that seem to deviate from the \citetalias{Benitez-Llambay2017} model are actually well reproduced by it once we account for their higher-than-average concentrations. In reality, their \ion{H}{I} mass is enhanced by two combined factors: their highly concentrated core, and their denser gas environment.

We speculate that there is a physical motivation behind this pairing. Since halo concentration strongly correlates with the formation time \citep[see, e.g.,][and references therein]{Wechsler2002, Ludlow2016}, these specific systems likely originated in dark matter environments that were slightly denser than the typical environments of mid-concentrated RELHICs ($c_{200}\sim 10$). This locally denser dark matter environment would force the halo to collapse earlier (leading to a higher concentration) while naturally dragging in more gas. Although these regions are not extreme (only $\sim 0.4$~dex denser than the baseline $\rm n_{H}=10^{-6}\,\mathrm{cm^{-3}}$), this could explain why high-concentration and denser gas environments are linked.

\section{Fits to synthetic RELHICs embedded in different environments}
\label{App:EnvironmentFits}

\begin{figure}[h]
    \centering
    \includegraphics[width=1\linewidth]{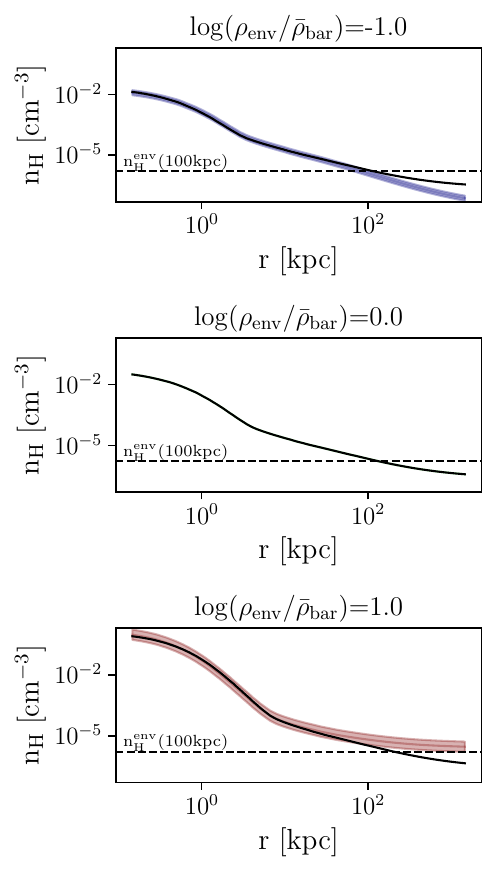}
       \caption{Synthetic gas density profiles (colored lines) for a fixed dark matter halo ($M_{200} = 10^{9.6}\,\rm M_\odot$, $c=10$) embedded in different local environments, compared with the best-fitting models (black lines) inferred via our nested sampling pipeline. The top panel features a RELHIC inhabiting an underdense environment ($\rho_{\rm env}/{\bar\rho}_{\rm bar} = 0.1$). The color coding is the same as in  Fig.~\ref{fig:Environment_theory}. The dashed horizontal lines indicate the expected density at 100 kpc according to~\citetalias{Benitez-Llambay2017} model. The profile drops below the model prediction at large radii, leading to an underestimated halo mass.  The middle panel shows a RELHIC in an environment at the cosmic mean density ($\rho_{\rm env}/{\bar\rho}_{\rm bar} = 1.0$). The recovery of the profile is accurate in this case. Finally, the bottom panel shows an overdense environment ($\rho_{\rm env}/\bar\rho_{\rm bar} = 10$). The high external pressure in this case elevates the gas density everywhere, causing the model to overestimate the halo mass to compensate.}
    \label{fig:Environment_theory_3profiles}
\end{figure}

In Sect. \ref{Subsec:EnvImpact} we demonstrated that the local environmental density systematically biases the recovery of RELHIC halo parameters. To illustrate exactly how the analytic model behaves under varying boundary conditions, Fig. \ref{fig:Environment_theory_3profiles} presents the individual best-fitting models for three representative synthetic profiles from our controlled experiment: an underdense environment ($\rho_{\rm env}/{\bar\rho}_{\rm bar} = 0.1$; top panel), the mean cosmic density ($\rho_{\rm env}/{\bar\rho}_{\rm bar} = 1.0$; middle panel), and an overdense environment ($\rho_{\rm env}/{\bar\rho}_{\rm bar} = 10.0$; bottom panel).
 When the background density matches the assumed cosmic mean boundary condition (middle panel), the \texttt{dynesty} pipeline perfectly recovers the true underlying halo mass ($M_{200} = 10^{9.6} \rm \ M_{\odot}$) and concentration ($c=10$), yielding an exact match to the synthetic profile at all radii.

However, in the overdense case (bottom panel), the intrinsic gas profile flattens at a systematically higher density in the outskirts. Forced to assume a rigid boundary condition at ${\bar\rho}_{\rm bar}$, the analytic model compensates for the globally elevated gas density by converging on an artificially high halo mass (and a correspondingly low concentration) to generate the gravitational force necessary to retain the excess of gas.

Conversely, in the underdense scenario (top panel), the profile drops below the model at large radii, causing our inference pipeline to underestimate the halo mass. Notably, despite the bias in the recovered parameters for non-average environments, the resulting parameters produce good fits to the synthetic data within the virial radius ($r \lesssim R_{200} \approx 30$ kpc). This exercise demonstrates the need to include the background density in the inference pipeline.

\section{Resolution-driven bias}
\label{App:Resolution}

\begin{figure}[h]
    \centering
    \includegraphics[width=1\linewidth]{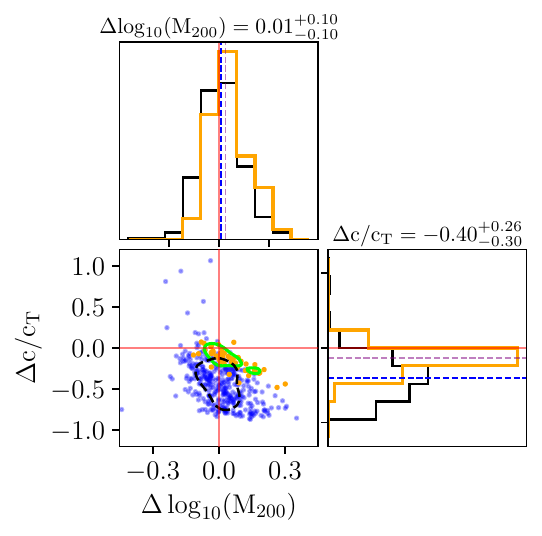}
    \caption{Distribution of the recovered halo parameters relative to the ground truth. Same as Fig.~\ref{fig:GasHist} but without the radial cut that imposes a minimum of 10 gas particles in the innermost radial bin.}
    \label{fig:GasHistFlat}
\end{figure}

As described in Sect.~\ref{Subsec:DensityProfiles}, we truncate the RELHIC density profiles below a minimum radius, $r_{\rm min}$, defined as the innermost radial bin containing at least ten gas particles. This radial cut is necessary to prevent the finite spatial resolution of our simulation from artificially biasing the parameter recovery. Figure~\ref{fig:GasHistFlat} illustrates the pipeline's performance when this cut is omitted. For intrinsically well-resolved systems (as defined in Sect.~\ref{Subsec:HaloParametersAll}), removing the cut has only a marginal impact on accuracy ($\Delta\log_{10} M_{200} = 0.03_{-0.07}^{+0.12}$, $\Delta c/c_{\rm T} = -0.14_{-0.16}^{+0.11}$). However, for poorly resolved systems, the limited numerical resolution dominates the error in the recovery parameters, driving a substantial bias of $\approx -40\%$ in the recovered concentration.

Although the radial cut introduced in Sect.~\ref{Subsec:DensityProfiles} improves the performance of the pipeline, a residual bias of $\sim -16\%$ persists in the recovered concentration. We ascribe this to limited resolution, given that both the bias and its dispersion drop significantly for well-resolved halos.

\begin{figure}[h]
    \centering
    \includegraphics[width=1\linewidth]{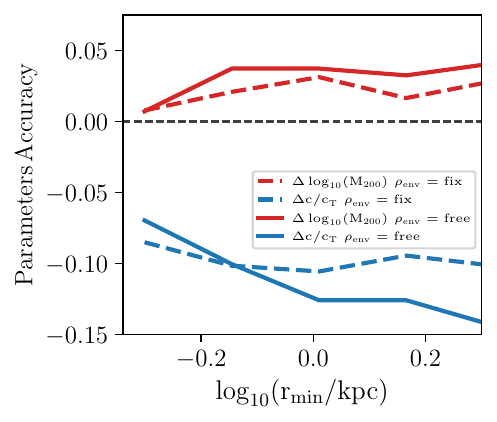}
    \caption{Accuracy of the recovered virial mass ($\Delta \log_{10} (M_{200})$) (red) and concentration ($\Delta c/c_{\rm T}$) (blue) as a function of the minimum radius $r_{\rm min}$ allowed in the subsample. Dotted and solid lines represent the pipeline performance with the environmental density fixed or treated as a free parameter, respectively. This exercise illustrates the significant reduction in concentration bias as a stricter resolution threshold is applied ($r_{\rm min} \leq 1$~kpc)}
    \label{fig:AccuracyVsr_min}
\end{figure}

While Sect.~\ref{Subsec:HaloParametersAll} restricted its analysis to a single well-resolved subsample ($r_{\rm min} \leq 1$~kpc), Fig.~\ref{fig:AccuracyVsr_min} systematically explores how these parameter biases depend on the adopted $r_{\rm min}$ threshold. The dotted and solid lines indicate the accuracy of the pipeline when the environmental density is fixed or treated as a free parameter, respectively. 

Further supporting our interpretation, we found that imposing increasingly strict radial cuts steadily mitigates the concentration bias. For the most conservatively resolved systems, the absolute concentration bias drops to $\approx 5\%$. Interestingly, this improvement is most pronounced when the environmental density is left as a free parameter, demonstrating that resolving the inner profile is essential for accurately disentangling the true halo concentration from boundary pressure effects.

\section{Decoupling environmental density}
\label{App:DecouplingDensity}

\begin{figure}
    \centering
    \includegraphics[width=\columnwidth]{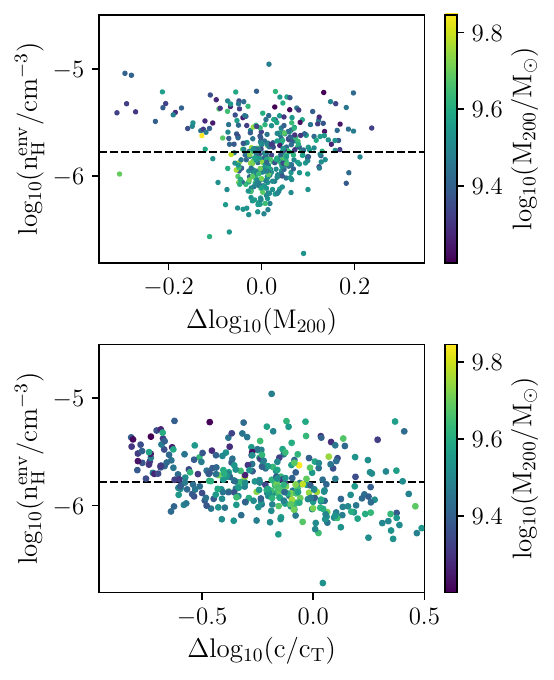}
    \caption{Residuals of the inferred halo parameters when treating the background environmental density as a free parameter in the inference pipeline. The top panel displays the residuals of the halo mass, $\Delta\log_{10} (M_{200})$, and the bottom panel displays the residuals of the concentration, $\Delta\log_{10}(c/c_{\rm T})$, plotted against the local environmental gas density, $\log_{10}(n_{\rm H}^{\rm env})$. The data points are colored by their true halo mass, $\log_{10}(M_{200}/{\rm M}_{\odot})$. While letting the background density vary successfully eliminates the correlation between the inference residuals and the environment for both parameters seen in Fig.~\ref{Fig:FigEnvironmentDensityAll}, a systematic absolute offset remains in the concentration estimates, which is likely driven by finite spatial resolution in the central regions of the simulated halos, as discussed in Appendix~\ref{App:Resolution}.}
    \label{Fig:app_rho_free}
\end{figure}

Figure~\ref{Fig:app_rho_free} shows that by allowing our inference pipeline to recover the background density, the correlation between the outer density of RELHICs and the residuals of the inference of the halo parameters disappears, further supporting the conclusion that fixing the background density artificially ties the inferred halo properties to their local environment. As illustrated in the top panel, treating the background density as a free parameter successfully mitigates the environmental trend for the halo mass, leaving the residuals ($\Delta\log_{10} (M_{200})$) distributed around zero across the probed range of environmental gas densities ($n_{\rm H}^{\rm env}$). Similarly, the bottom panel demonstrates that the systematic tilt between the concentration residuals ($\Delta\log_{10}(c/c_{\rm T})$) and the environment seen in Fig.~\ref{Fig:FigEnvironmentDensityAll} is effectively removed. However, while decoupling the background density eliminates the environmental dependence, it does not yield perfectly unbiased absolute estimates for the concentration. The overall offset visible in the bottom panel suggests an underlying systematic effect remains. As discussed above, we speculate that this lingering absolute bias is fundamentally limited by the spatial resolution of the simulations, which prevents the pipeline from perfectly resolving the innermost halo structure even when the outer boundary conditions are modeled accurately.

\end{appendix}

\end{document}